\documentclass[floatfix,a4paper,tightenlines,amsmath,amssymb,nofootinbib,showpacs,notitlepage,showkeys,prd,twocolumn,10pt,aps,superscriptaddress]{revtex4-2}

\usepackage[utf8]{inputenc}
\usepackage{amsmath}
\usepackage{amssymb}
\usepackage{graphicx}
\usepackage{hyperref}
\usepackage{float}
\usepackage{placeins}
\usepackage[export]{adjustbox}
\usepackage[usenames,dvipsnames,svgnames,table]{xcolor}

\newcommand{\de}{\delta}
\newcommand{\eref}[1]{Eq. (\ref{#1})}
\newcommand{\fref}[1]{Fig. \ref{#1}}
\newcommand{\tref}[1]{Tab.~\ref{#1}}
\newcommand{\nnnl}{\nonumber\\}	

\allowdisplaybreaks 

\begin{document}

\title{Yang-Mills propagators in linear covariant gauges from Nielsen identities}

\author{Martin Napetschnig}
\email{martin.napetschnig@edu.uni-graz.at}
\affiliation{Institute of Physics, University of Graz, NAWI Graz, Universit\"atsplatz 5, 8010 Graz, Austria}

\author{Reinhard~Alkofer}
\email{reinhard.alkofer@uni-graz.at}
\affiliation{Institute of Physics, University of Graz, NAWI Graz, Universit\"atsplatz 5, 8010 Graz, Austria}

\author{Markus Q.~Huber}
\email{markus.huber@physik.jlug.de}
\affiliation{Institut f\"ur Theoretische Physik, Justus-Liebig-Universit\"at Giessen, 35392 Giessen, Germany}

\author{Jan M. Pawlowski}
\email{j.pawlowski@thphys.uni-heidelberg.de}
\affiliation{Institute for Theoretical Physics, Universit\"at
  Heidelberg, Philosophenweg 12, D-69120 Germany}
\affiliation{ExtreMe Matter Institute EMMI, GSI, Planckstr. 1,
  D-64291 Darmstadt, Germany}

\date{\today}

\begin{abstract}
We calculate gluon and ghost propagators in Yang-Mills theory in linear covariant gauges. 
To that end, we utilise Nielsen identities with Landau gauge propagators and vertices as the starting point. 
We present and discuss numerical results for the gluon and ghost propagators for values of the gauge parameter 
$0<\xi \le 5$. Extrapolating the propagators to $\xi \to \infty $ we find the expected qualitative behavior.
We provide arguments
that our results are quantitatively reliable at least for values $\xi\lesssim 1/2$ of the gauge fixing parameter.
It is shown that the correlation functions, and in particular the ghost propagator, change significantly with increasing gauge 
parameter.
In turn, the ghost-gluon running coupling as well as the position of the zero crossing of the Schwinger function of the gluon propagator remain within the uncertainties of our calculation unchanged. 

\end{abstract}



\maketitle

\section{Introduction}

In the past decades, functional approaches such as Dyson-Schwinger equations (DSEs) or functional renormalization group (fRG) equations have very successfully contributed to understanding many phenomena in quantum chromodynamics (QCD), ranging from the hadron resonance spectrum to the phase structure of QCD at nonvanishing temperatures and densities. The majority of the respective investigations have been carried out in the Landau gauge due to the technical as well as conceptual advantages this gauge provides.

Evidently, obtaining via such approaches results for physical observables requires truncations to the full hierarchy of coupled functional equations, typically chosen to be of a given order in a systematic approximation scheme such as the vertex expansion. This calls for checks of the systematic errors of the respective results.
For example, the gauge independence of the computed observables would be a very powerful self-consistency check.
Although demonstrating generic gauge independence is likely beyond reach within functional approaches, the test of a reasonably accurate independence of gauge-invariant quantities when varying the gauge parameter within a given class of gauges would provide a convincing (but also costly) verification of the employed truncations.

In a first step towards such a  self-consistency check, we study in this work the propagators of elementary Yang-Mills fields in the linear covariant gauges thus extending \cite{Huber:2015ria}. In addition, our investigation may serve for corroborating the current state of the art of functional studies in the Landau gauge, for the respective recent Landau gauge DSE results for propagators and vertex functions see, e.g., \cite{Huber:2020keu,Huber:2018ned},
and for recent quantitative fRG results \cite{Cyrol:2016tym,Dupuis:2020fhh}. These Yang-Mills results within the Landau gauge, in particular for the ghost and gluon propagators, match quantitatively the 
respective available lattice results, see, e.g., \cite{Cucchieri:2007md, Cucchieri:2008fc, Bogolubsky:2007ud, Bogolubsky:2009dc}. 
Detailed discussions of Yang-Mills correlation functions in the Landau gauge as well as more results on them can be found, for instance, in Refs. \cite{Alkofer:2000wg,Fischer:2006ub,Fischer:2008uz, Aguilar:2008xm, Dudal:2008sp, Dudal:2011gd, Maas:2011se, Boucaud:2011ug, Quandt:2013wna, Cyrol:2016tym, Aguilar:2019kxz, Huber:2018ned, Huber:2020keu, Pelaez:2021tpq,Dupuis:2020fhh} and references therein.

Herein, we extend previous studies \cite{Aguilar:2015nqa,Huber:2015ria} and compute ghost and gluon propagators from Nielsen identities (NIs). Further results on linear covariant gauges from functional methods can be found, e.g., in \cite{Breckenridge:1994gs, Alkofer:2003jr, Aguilar:2007nf, Huber:2009wh, Aguilar:2015nqa, Huber:2015ria, Aguilar:2016ock, DeMeerleer:2019kmh}, from the (refined) Gribov-Zwanziger framework in \cite{Sobreiro:2005vn,Capri:2015ixa,Capri:2015nzw,Capri:2016gut}, from variational methods in \cite{Siringo:2014lva, Siringo:2015gia, Siringo:2018uho}, and from lattice methods in \cite{Cucchieri:2009kk, Cucchieri:2011pp, Bicudo:2015rma, Cucchieri:2018leo, Cucchieri:2018doy}; see also the respective part in the recent review \cite{Huber:2018ned} and references therein.

The NIs describe the dependence of correlation functions on the gauge fixing parameter in terms of a differential equation of the effective action w.r.t.\ the gauge fixing parameter, see, e.g., \cite{Breckenridge:1994gs, Aguilar:2015nqa, DeMeerleer:2019kmh}. The resulting equations for the correlations functions can be integrated from the Landau gauge to any linear covariant gauge, and we are going to report on an investigation in which we used the quantitative DSE results for Landau gauge correlation functions from Ref.\ \cite{Huber:2020keu} as a starting point for such an integration of a set of coupled differential equations.   

Moreover, NIs have also been derived for other families of gauges. Often, these families
 go under the name interpolating gauges. Examples include interpolating gauges between the Landau gauge and the 
Coulomb gauge \cite{Cucchieri:2007uj, Andrasi:2021qrw}, the Landau gauge and the maximally Abelian gauge 
\cite{Dudal:2004rx, Huber:2009wh, Huber:2010ne}, the linear covariant gauges, the Coulomb gauge and the maximally 
Abelian gauge \cite{Capri:2005zj, Capri:2006bj}, and the linear covariant gauges, the maximally Abelian gauge and 
the Curci-Ferrari gauge \cite{Dudal:2005zr}.

This article is structured as follows. In the next section, we introduce the correlation functions and their functional equations.
In Sec.~\ref{sec:setup}, the setup including truncations is presented and discussed.
Sec.~\ref{sec:results} contains the results, and we conclude in Sec.~\ref{sec:summary} with a summary.
Several appendices contain technicalities including discussions of the RG and UV properties of the equations and the model parameter dependence of the solution.

\section{Correlation functions and their Nielsen identities}
\label{sec:corrFuncs_NIs}

As usual in functional approaches it is assumed that a Wick rotation to Euclidean space has been performed.
The Lagrangian density of Yang-Mills theory in linear covariant gauges is then given by
\begin{align}\label{eq:Lagrangian}
 \mathcal{L}&=\mathcal{L}_\text{YM}+\mathcal{L}_\text{gf}\,,
\end{align}
with
\begin{align}\nonumber 
 \mathcal{L}_\text{YM}&=\frac{1}{4} F_{\mu \nu }^aF^a_{\mu \nu}\,,\\[1ex]\nonumber 
 F^{a}_{\mu\nu}&=\partial_\mu A_\nu^a-\partial_\nu A_\mu^a-g\,f^{abc}A_\mu^b A_\nu^c\,,\\[1ex]
 \mathcal{L}_\text{gf}&=s\left(\overline{c}^a \partial_\mu A_\mu^a-i \frac{\xi}{2}\overline{c}^a b^a\right)\, .
\end{align}
The fields are the gluon field $A_\mu^a$, the ghost field $c^a$, the anti-ghost field $\overline{c}^a$ and 
the Nakanishi-Lautrup field $b^a$. 
The Nakanishi-Lautrup fields are introduced via the Becchi-Rouet-Stora-Tyutin (BRST) transformation \cite{Becchi:1975nq,Tyutin:1975qk}, denoted by $s$, and given by
\begin{align}\nonumber 
s\,A_\mu^a&=-D_\mu^{ab} c^b, \\[1ex]\nonumber 
s\,c^a&=-\frac{1}{2} g\,f^{abc} c^b c^c, \\[1ex]\nonumber 
s\,\bar{c}^a&=i \, b^a,\\[1ex]
s\,b^a&=0\,,
\label{eq:BRST}\end{align}
where 
\begin{align}
	D_\mu^{ab}&=\delta^{ab}\partial_\mu+g\,f^{abc} A_\mu^c \, ,
\end{align}
is the covariant derivative in the adjoint representation. 
Then, the gauge fixing part of the Lagrangian reads explicitly
\begin{align}
\mathcal{L}_\text{gf}&=i\,b^a\,\partial_\mu A^a_\mu-\overline{c}^a\, \partial_\mu (-D^{ab}_\mu c^b)
-\frac{i}{2}\chi\, \overline{c}^a\, b^a\,,
\end{align}
where $\chi=s\,\xi$ was introduced as the BRST transformation of the gauge fixing parameter $\xi$.

The BRST transformations are nonlinear for the gluon and ghost fields.
We are going to work with the Batalin-Vilkovsky (BV) or antifield formalism, see, e.g., \cite{Batalin:1981jr,Batalin:1983wj,Batalin:1984jr,Fuster:2005eg}, and introduce sources, called antifields, for them, 
\begin{align}\nonumber 
\mathcal{L}_{\text{BV}} &= -A_\mu^{*a} s \, A^a_\mu - c^{a*} s \, c^a\\[1ex]
&=A_\mu^{*a} D^{ab}_\mu c^b + \frac{g}{2} f^{abc} c^{*a} c^b c^c.
\label{eq:brst_var}\end{align}
In $\mathcal{L}_\text{BV}$, additional vertices appear that contain an antifield $A^*$ or $c^*$ and are hence called antifield vertices.

In the Landau gauge, it is sufficient to consider the completely transverse part of the dressing functions as they form a closed system \cite{Fischer:2008uz, Dupuis:2020fhh}, and we split the gluon propagator into a transverse and longitudinal part written as
\begin{align}\nonumber 
 D^{ab}_{\mu\nu}(p)&=\de^{ab}D_{\mu\nu}(p)=\de^{ab}(D^T_{\mu\nu}(p)+D^L_{\mu\nu}(p))\,,\\[1ex]\nonumber 
 D^T_{\mu\nu}(p)&=P_{\mu\nu}(p)\frac{Z(p^2)}{p^2}\,,\\[1ex]
 D^L_{\mu\nu}(p)&=\frac{p_\mu p_\nu}{p^2}\frac{Z^L(p^2)}{p^2}\,, 
\end{align}
where $P_{\mu\nu}(p)=g_{\mu\nu}-p_\mu p_\nu /p^2$ is the transverse projection operator. The Slavnov-Taylor identity
(STI) for the gluon propagator enforces its longitudinal dressing function to be constant, $Z^L(p^2)=\xi$, i.e., all quantum corrections are transverse. Correspondingly, in the Landau gauge, $\xi=0$, the gluon propagator is proportional to a transverse projection operator. For $\xi>0$, this is no longer true and also the trivial longitudinal part enters.

The ghost has only one dressing function, 
\begin{align}
 D^{ab}_G(p)=-\de^{ab}\frac{G(p^2)}{p^2}\,.
\end{align}
The ghost-gluon vertex can be conveniently split into a transverse and a longitudinal part, 
\begin{align}\nonumber 
\Gamma_{A^a_\mu \bar c^b c^c}(k;p,q)=
&\, -i\,g\,f^{abc}\Bigl(D^{A\bar c c,T}(k;p,q)P_{\mu\nu}(k)p_\nu\\[1ex] 
& \hspace{1cm}+D^{A\bar c c,L}(k;p,q)k_\mu\Bigr)\,.
\end{align}
We use a compact notation where the subscripts denote the fields with indices corresponding to the momentum arguments.
The tree-level expression is, respectively, $D^{A\bar c c,T}(k;p,q)=1$ and $D^{A\bar c c,L}(k;p,q)=p\cdot k/k^2$.
For the three-gluon vertex we only use a dressed tree-level tensor, thereby neglecting components which are sub-leading
\cite{Eichmann:2014xya},
\begin{align}
 &\Gamma_{A^a_\mu A^b_\nu A^c_\rho}(p,q,r)=i\,g\,f^{abc}C^{AAA}(p,q,r)\nnnl
 &\quad\times\Big[(p-q)_\rho g_{\mu\nu}+(q-r)_\mu g_{\nu\rho}+(r-p)_\nu g_{\mu\rho}\Big]\,.
\end{align}
The NI encodes the dependence of the effective action $\Gamma$ on the gauge fixing parameter $\xi$ and reads 
\begin{align}\label{eq:STI_explicit}
    \frac{\partial \Gamma}{\partial \xi} \chi =\int d x \left( \frac{\delta \Gamma}{\delta A_\mu^a} \frac{\delta \Gamma}{\delta A_\mu^{*a}} + \frac{ \delta\Gamma }{\delta c^a} \frac{\delta \Gamma}{\delta c^{*a}} + i \frac{\delta \Gamma}{\delta \bar{c}^a} b^a \right)\,.
\end{align}
The right hand side of (\ref{eq:STI_explicit}) follows from the BRST invariance of the effective action. For the final form we differentiate (\ref{eq:STI_explicit}) with respect to $\chi$ and set $\chi=0$.
This leads to a master equation for the NIs:
\begin{align}\nonumber 
 \left. \frac{\partial \Gamma}{\partial \xi}\right\vert_{\chi = 0} = \int dx \left. \left( \frac{ \partial\delta \Gamma }{\partial \chi \delta A_{\mu}^{a} } \frac{\delta \Gamma}{\delta A^{* a}_{\mu}} + \frac{\delta \Gamma}{\delta A_{\mu}^{a}} \frac{ \partial \delta \Gamma}{ \partial \chi \delta A^{* a}_{\mu}}  \right. \right.\\[2ex]
\left. \left.  
+\frac{\partial \delta \Gamma }{ \partial \chi \delta c^{ a} } \frac{\delta \Gamma}{\delta c^{^*a}} 
-\frac{\delta \Gamma}{\delta{c^a}} \frac{ \partial \delta \Gamma}{ \partial \chi \delta c^{*a} } +i b^a \frac{\partial \delta \Gamma }{ \partial \chi \delta \bar{c}^a}
 \right)\right\vert_{\chi = 0}\,.
\label{eq:NIfinal}\end{align}
NIs for correlation functions can now be obtained by applying further field derivatives to (\ref{eq:NIfinal}).
For related work on Nielsen identities see \cite{Binosi:2002ez,Binosi:2005yk}.

For the propagators, the NIs in the form (\ref{eq:NIfinal}) used here were derived in Ref.~\cite{Breckenridge:1994gs}, where also a perturbative one-loop analysis was done, and we refer to this reference for further details. The NIs for the ghost and gluon
propagators read in Euclidean momentum space:
\begin{subequations}
\label{eq:NIs}
\begin{align}
 \partial_\xi \Gamma_{\bar cc}(p^2)= &\,\bigg[i \frac{p_\mu}{p^2} \Gamma_{\overline{c} \chi A_\mu}(p,0,-p) \\[1ex]\nonumber 
 & \hspace{.4cm} +\Gamma_{c \chi c^*}(p,0,-p) \bigg] \Gamma_{\bar cc}(p^2)\,,\\
\partial_\xi \Gamma_{A_\mu A_\nu}(p^2) = &\, 2 \Gamma_{A^*_{\rho} \chi A_\mu}(p,0,-p)\Gamma_{A_\nu A_\rho}(p^2) \,, 
\end{align}
\end{subequations}
where we have suppressed the color indices. For the numerical solution of (\ref{eq:NIs}) we use approximations for the vertices $\Gamma_{\overline{c} \chi A}$, $\Gamma_{c\chi c^*}$ and $\Gamma_{A\chi A^*}$ which will be discussed in detail in Sec.~\ref{sec:setup}.

For the ghost-gluon, the three-gluon and the four-gluon vertices the NIs read
\begin{widetext}
\begin{align}\nonumber 
    \partial_\xi \: \Gamma_{A_\mu^a \overline{c}^b c^c} &= -i\,\frac{p_\rho}{p^2}
    \Gamma_{ \chi\overline{c}^b A^d_\rho A^a_\mu}\Gamma_{\overline{c}^d c^b}
    +\Gamma_{ \chi\overline{c}^b A^d_\rho}\Gamma_{c^c A^{*d}_\rho A^a_\mu}
    +\Gamma_{ c^c\chi c^{*d}}\Gamma_{A^a_\mu \overline{c}^b c^d }\\[1ex]\nonumber 
        &\quad +\Gamma_{ \overline{c}^bc^d}\Gamma_{A^a_\mu c^c \chi c^{*d}}
    +\Gamma_{ A^a_\mu A^d_\rho}\Gamma_{\overline{c}^b c^c \chi A^{*d}_\rho}
    +\Gamma_{ A^d_\rho
    \overline{c}^b c^c}\Gamma_{\chi A^a_\mu A^{*d}_\rho}\,,\\[2ex]\nonumber 
    \partial_\xi \: \Gamma_{A_\mu^a A_\nu^b A_\rho^c} &= 
    \Gamma_{ \chi A_\sigma^{*d} A_\mu^a A_\nu^b } \Gamma_{A_\rho^c A_\sigma^d}
    + \Gamma_{\chi A_\sigma^{*d} A_\mu^a} \Gamma_{ A_\nu^b A_\rho^c A_\sigma^d}+\text{permutations}\,,\\[2ex]
    \partial_\xi \: \Gamma_{A_\mu^a A_\nu^b A_\rho^c A_\sigma^d} &=
    \Gamma_{\chi A^{*e}_\tau A_\mu^a A_\nu^b A_\rho^c} \Gamma_{A_\sigma^d A_\tau^e} 
    + \Gamma_{\chi A^{*e}_\tau A_\mu^a A_\nu^b } \Gamma_{A_\rho^c A_\sigma^d A_\tau^e}  
    + \Gamma_{\chi A^{*e}_\tau A_\mu^a  } \Gamma_{A_\nu^b A_\rho^c A_\sigma^d A_\tau^e}  +\text{permutations}\,.
\end{align}
\end{widetext}

\section{Truncation and input}
\label{sec:setup}

\begin{figure}[tb]
 \includegraphics[width=0.49\textwidth]{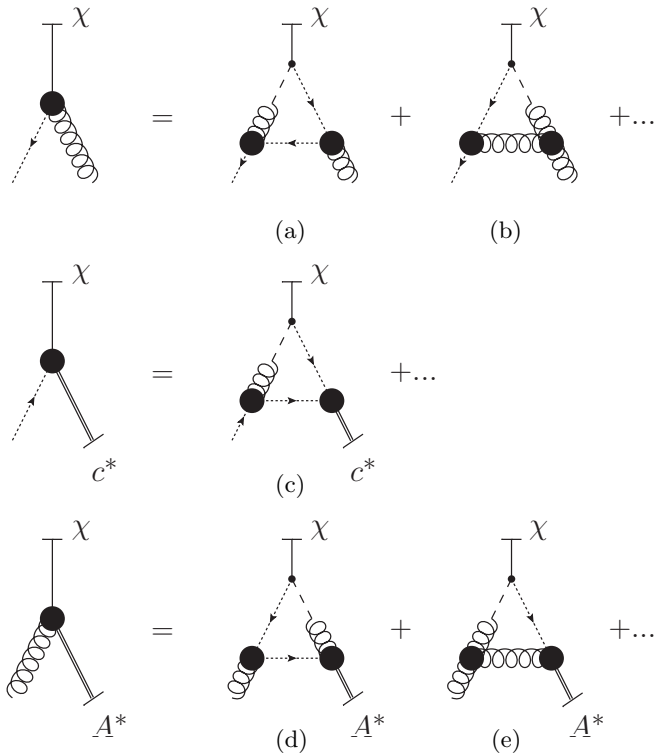}\\
 \vskip-70mm \hskip20mm (a) \hskip23mm (b)\\
 \vskip30mm
 \hskip20mm (c) \hskip23mm \phantom{(b)}\\
 \vskip30mm
 \hskip20mm (d) \hskip23mm (e)\\
 \caption{Skeleton expansions for $\Gamma_{\overline{c}\chi A_\mu}$, $\Gamma_{c\chi c^*}$, and $\Gamma_{A\chi A^*}$ (top to bottom).
 Wiggly lines denote gluons and dotted ones ghosts. Wiggly-dashed ones represent a mixed gluon-Nakanishi Lautrup field propagator.
 The antifields $A^*$ and $c^*$ as well as $\chi$ are indicated explicitly.
 Propagators are all dressed, small/large dots represent bare/dressed vertices.}
 \label{fig:skelExp}
\end{figure}

The NIs are exact functional equations, and we cannot solve them without approximations. 
For the propagator equations we need $\Gamma_{\overline{c} \chi A_\mu}(p,0,-p)$, $\Gamma_{c \chi c^*}(p,0,-p)$ and $\Gamma_{A_\mu \chi A^*_\nu}(p,0,-p)$. We follow Ref.~\cite{Aguilar:2015nqa} and calculate them from the first order in a skeleton expansion, also called dressed-loop expansion, shown in \fref{fig:skelExp}. 
An additional approximation in Ref.~\cite{Aguilar:2015nqa} was the use of bare vertices. In this work, however, we keep the vertices dressed.
The resulting expressions $K^{(i)}$ arising on the right-hand side of \eref{eq:NIs} correspond to the loop diagrams 
$i=a,b,c,d,e$ in \fref{fig:skelExp} and are provided in Appendix~\ref{app:K(i)}. 

Using these expressions and switching from the two-point functions $\Gamma_{AA}$ and $\Gamma_{\bar cc}$ to the dressing function $Z$ and $G$, respectively, the propagator NIs can be written as
\begin{align}\nonumber 
 \partial_\xi \ln G(p^2)&= K^{(a)}+K^{(b)}+K^{(c)}, \\[1ex]
\partial_\xi \ln Z(p^2) &=  K^{(d)}+K^{(e_1)}+K^{(e_2)} \, ,
\label{eq:NI_K_G}\end{align}
where the $i=e$ contribution has been split into two terms for convenience, cf., Eqs.~(\ref{eq:Ks}).

\begin{figure*}[tb]
 \includegraphics[width=0.49\textwidth]{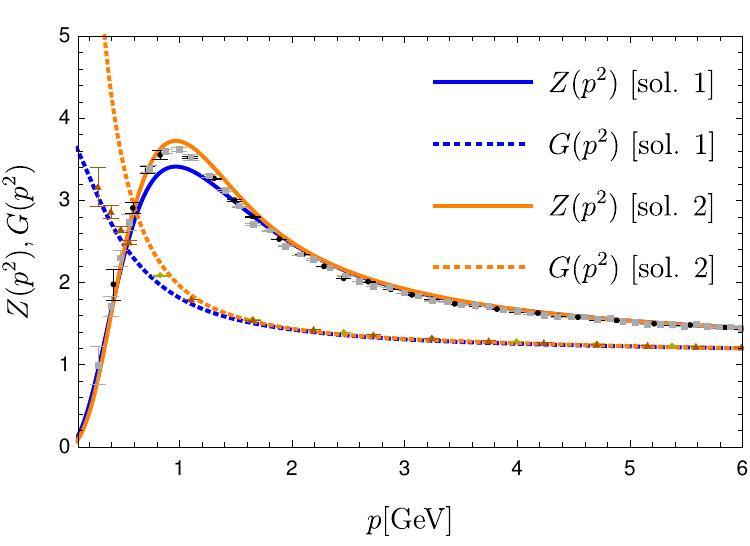}\hfill
 \includegraphics[width=0.49\textwidth]{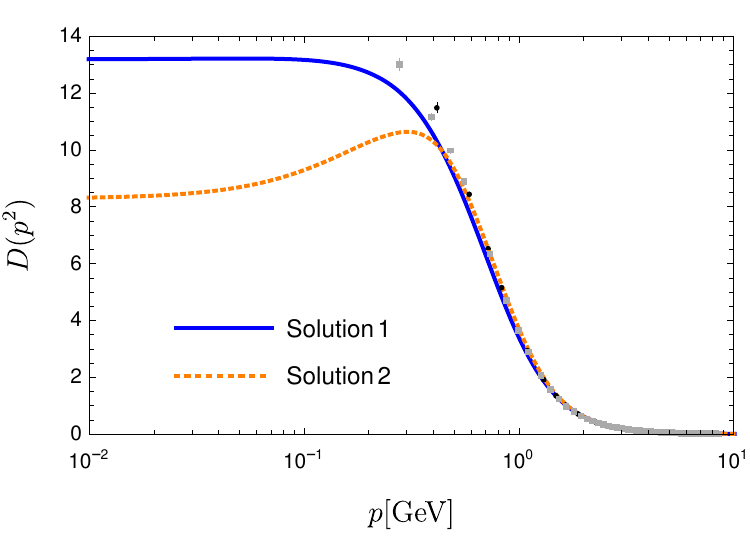}
 \caption{Ghost and gluon dressing functions (left) and gluon propagator (right) from lattice \cite{Sternbeck:2006rd} and DSE calculations \cite{Huber:2020keu} in the Landau gauge.}
 \label{fig:propsLG}
\end{figure*}

\begin{figure*}[tb]
 \includegraphics[width=0.49\textwidth]{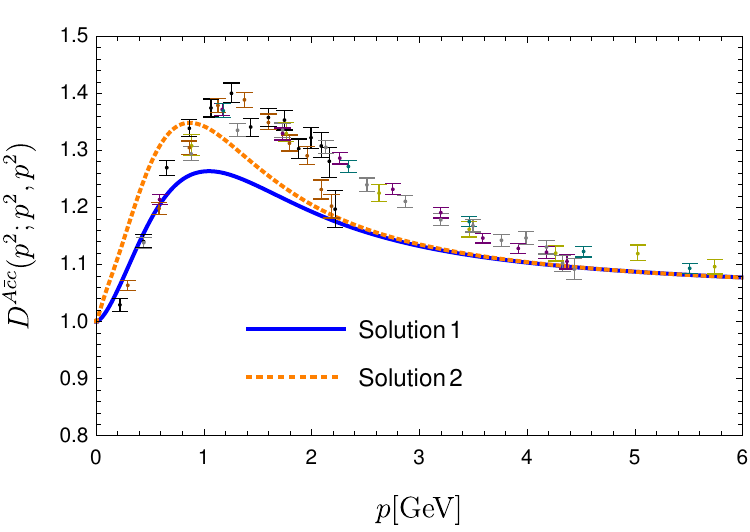}\hfill
 \includegraphics[width=0.49\textwidth]{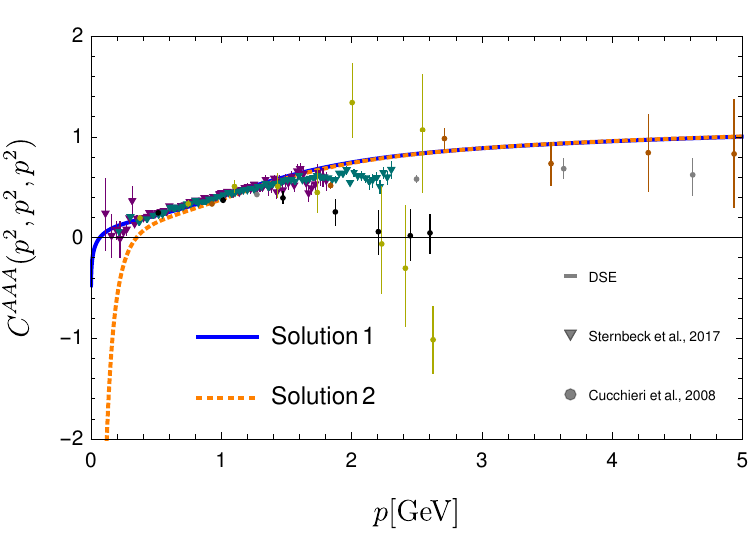}
 \caption{Ghost-gluon (left) and three-gluon vertices (right) in Landau gauge from lattice \cite{Maas:2019ggf,Cucchieri:2008qm,Sternbeck:2017ntv} and DSE calculations \cite{Huber:2020keu} in the Landau gauge.
 For more lattice results on the three-gluon vertex see, e.g., \cite{Athenodorou:2016oyh,Duarte:2016ieu,Aguilar:2021lke}.}
 \label{fig:vertsLG}
\end{figure*}

\begin{figure*}[tb]
 \includegraphics[width=0.48\textwidth]{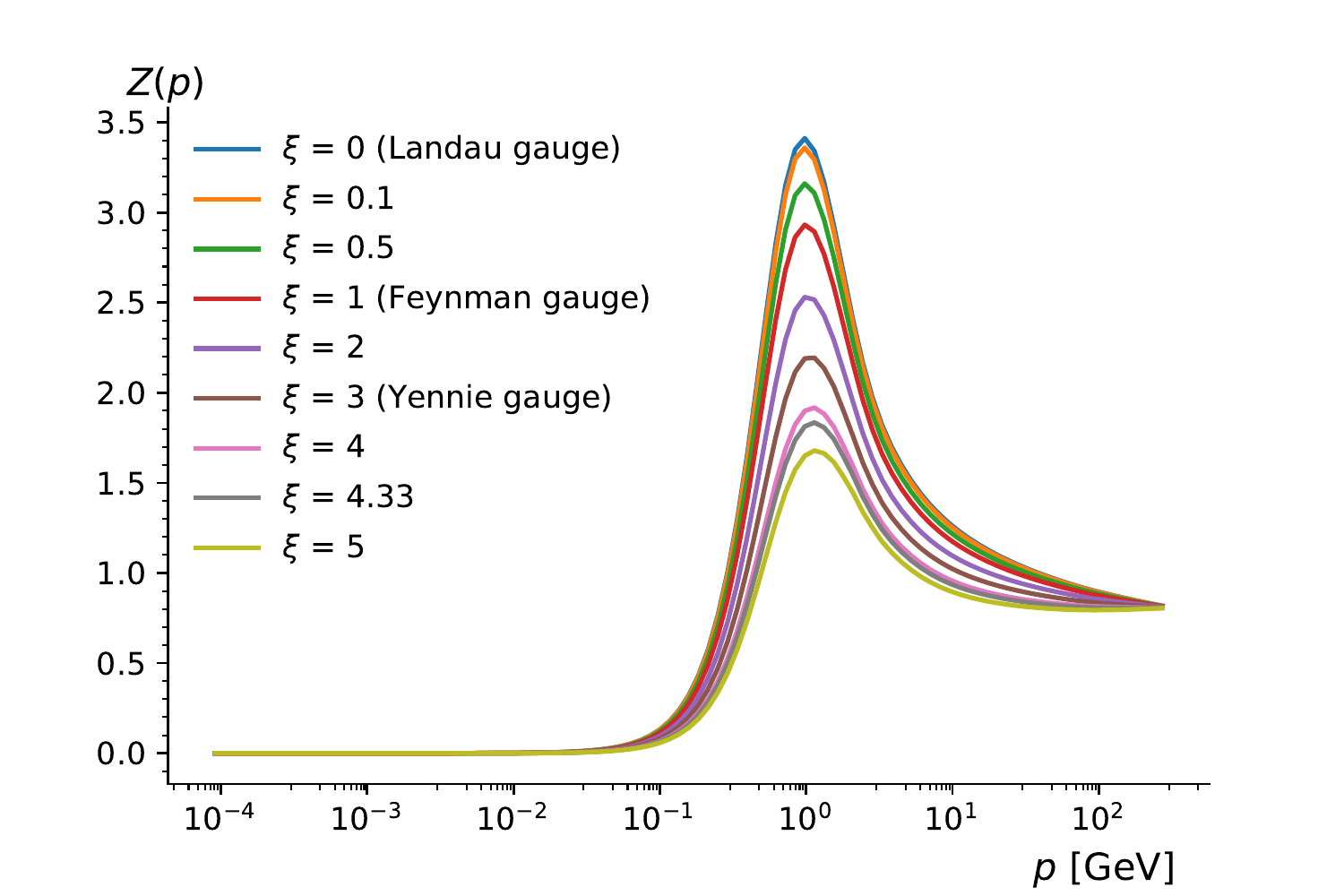}\hfill
 \includegraphics[width=0.48\textwidth]{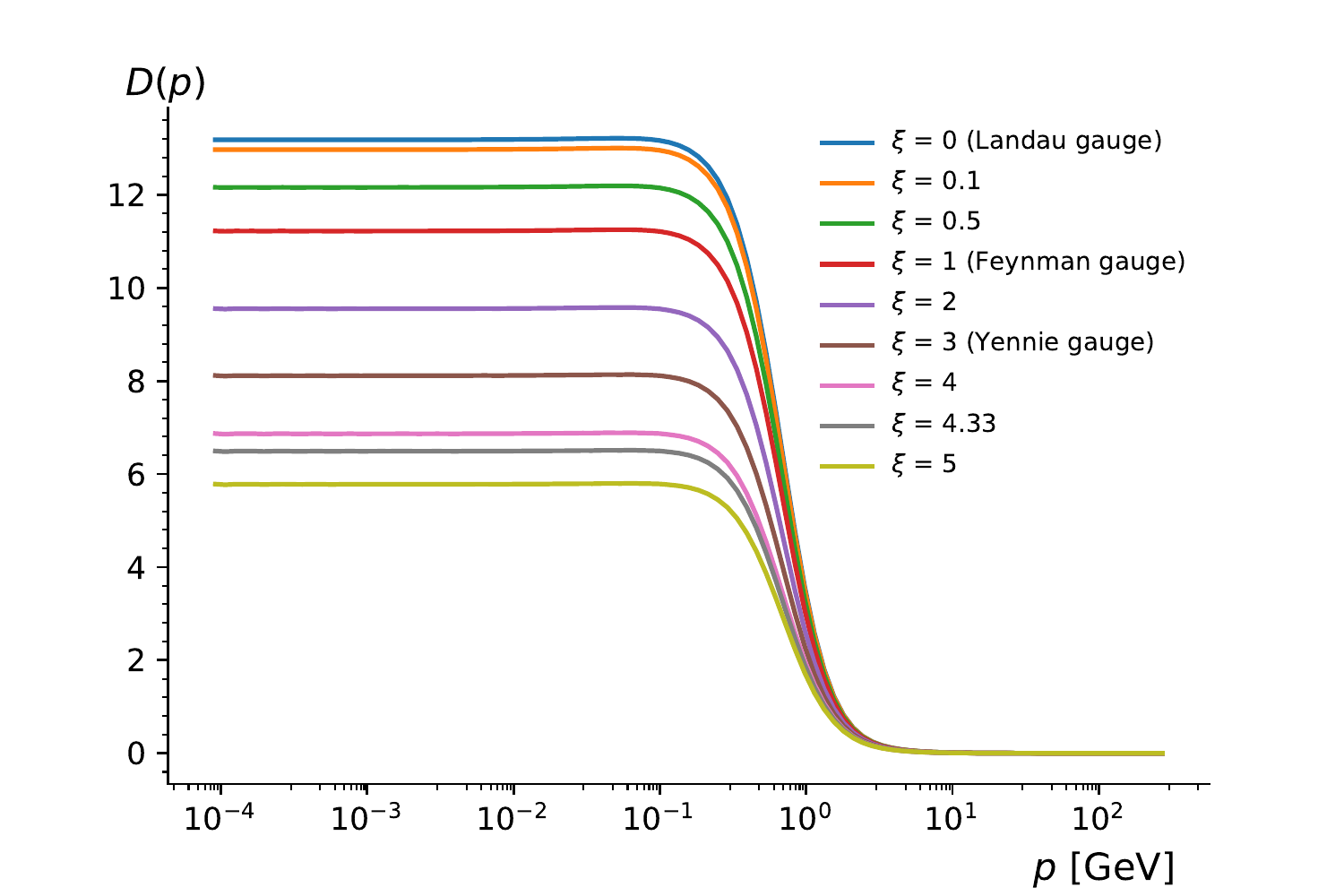}
 \caption{The gluon dressing function (left) and propagator (right) for various $\xi$ including the starting point $\xi=0$.}
 \label{fig:res_gl}
\end{figure*}

From the loop diagrams, we can directly confirm that the NIs transform correctly under a change of the renormalization scale $\mu_1\to\mu_2$. The NIs only depend on renormalized correlation functions and couplings ${\cal O}_i(p,\mu)$ that schematically change as 
\begin{align}\label{eq:RGrescalings}
{\cal O}_i(p,\mu_1 ) \to \frac{Z_{{\cal O}_i}(\Lambda,\mu_1)}{Z_{{\cal O}_i}(\Lambda,\mu_2)}{\cal O}_i(p,\mu_2) \,,
\end{align}
where $Z_{{\cal O}_i}(\Lambda,\mu)$ is the renormalization factor of the respective correlation function or coupling ${\cal O}_i$. The 
	explicit forms are deferred to Appendix~\ref{app:RGrescalings}. 

We exemplify the consistency of the left- and right-hand sides of the NIs under a change of the renormalization scale with diagram (a).
Under a change of the renormalization scale $\mu\rightarrow \nu$, diagram (a) transforms, upon using \ref{eq:STIs}, as
\begin{align}
 \frac{Z_g(\Lambda,\nu)^2 \widetilde Z_3(\Lambda,\nu)^2 \widetilde Z_1(\Lambda,\mu)^2 }
 {Z_g(\Lambda,\mu)^2 \widetilde Z_3(\Lambda,\mu)^2 \widetilde Z_1(\Lambda,\nu)^2 }=\frac{Z_3(\Lambda,\mu)}{Z_3(\Lambda,\nu)}.
\end{align}
This is exactly as the $\xi$ derivative on the left-hand side of \eref{eq:NI_K_G} transforms.
For all other diagrams this analysis can be repeated and leads to the same result.

As starting value for integrating the NIs we use the Landau gauge, $\xi=0$, for which we have results for the propagators and vertices.
However, we also the need the vertices for $\xi>0$.
Based on the fact that the $\xi$ dependence for the propagators found on the lattice \cite{Cucchieri:2009kk,Cucchieri:2011pp,Bicudo:2015rma,Cucchieri:2018leo,Cucchieri:2018doy} is small, we adopt as working assumption that the ghost-gluon and three-gluon vertices deviate only little as well and use results for $\xi=0$ for all $\xi$.
Only in the UV we accommodate the correct $\xi$ dependence by modifying the anomalous running accordingly.
In addition, we approximate their longitudinal parts with the transversely projected ones due to the lack of concrete results for the former.
A final approximation consists in taking only a single kinematic configuration for each vertex.
For the three-gluon vertex, this is a good approximation due to its small angular dependence \cite{Blum:2014gna,Eichmann:2014xya,Huber:2020keu, Aguilar:2013vaa}.
The ghost-gluon vertex shows more angular dependence, which we neglect here, though. 
This is justified by the overall modest variation of the ghost-gluon vertex with respect to momenta.

The Landau gauge results \cite{Huber:2020keu,Huber:2020git} used as initial values for solving the NIs are shown in Figs.~\ref{fig:propsLG} and \ref{fig:vertsLG} in comparison to lattice results.
Specifically, we use a self-contained solution that possess several advantageous properties.
Among them are manifest gauge covariance expressed by the good agreement of different couplings in the perturbative
 regime and a unique treatment of quadratic divergences,
for details we refer to \cite{Huber:2020keu}.
We also fix the overall scale from these results which which was obtained by matching the maximum of the gluon dressing function to lattice results.

In the Landau gauge it is well studied that a family of different solutions can be obtained from functional equations 
which differ in their IR behavior \cite{vonSmekal:1997vx,vonSmekal:1997is,Zwanziger:2001kw,Lerche:2002ep,Pawlowski:2003hq,Fischer:2008uz,Alkofer:2008jy}.
These solutions are different only in the region below $2\,\text{GeV}$.
Most notably, the maximum in the gluon propagator changes.
To explore the existence of such solutions also beyond the Landau gauge we choose two different sets of solutions.
One of the solutions is the one that agrees best with lattice results and has only a shallow, hardly visible maximum in the gluon propagator.
As a second choice, we take a solution with a more pronounced maximum, see \fref{fig:propsLG}.

It remains to specify the models used for the vertices $\Gamma_{A^*Ac}$ and $\Gamma_{c^*cc}$.
When using bare vertices, we found that in the infrared (IR) individual loop diagrams can qualitatively modify the 
IR solution for the gluon propagator.
This is either resolved by cancellations between individual diagrams or by the IR behavior of the antifield vertices.
We explore the second option, as it is currently not clear if the first one can be realized in the truncation we use.

\begin{figure}[tb]
 \includegraphics[width=0.48\textwidth]{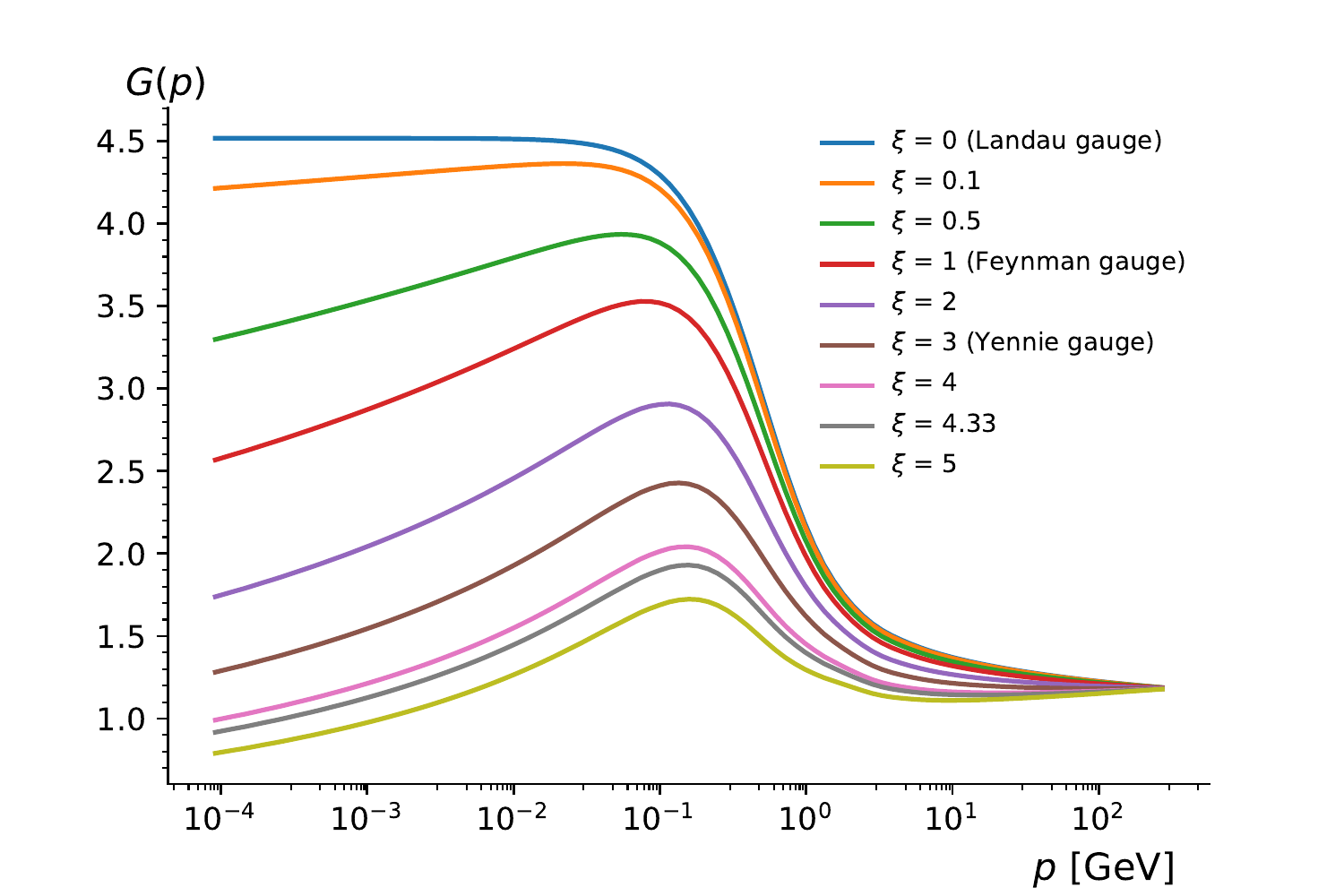}
 \caption{The ghost dressing function for various $\xi$ including the starting point $\xi=0$.}
 \label{fig:res_gh}
\end{figure}

\begin{figure}
 \includegraphics[width=0.48\textwidth]{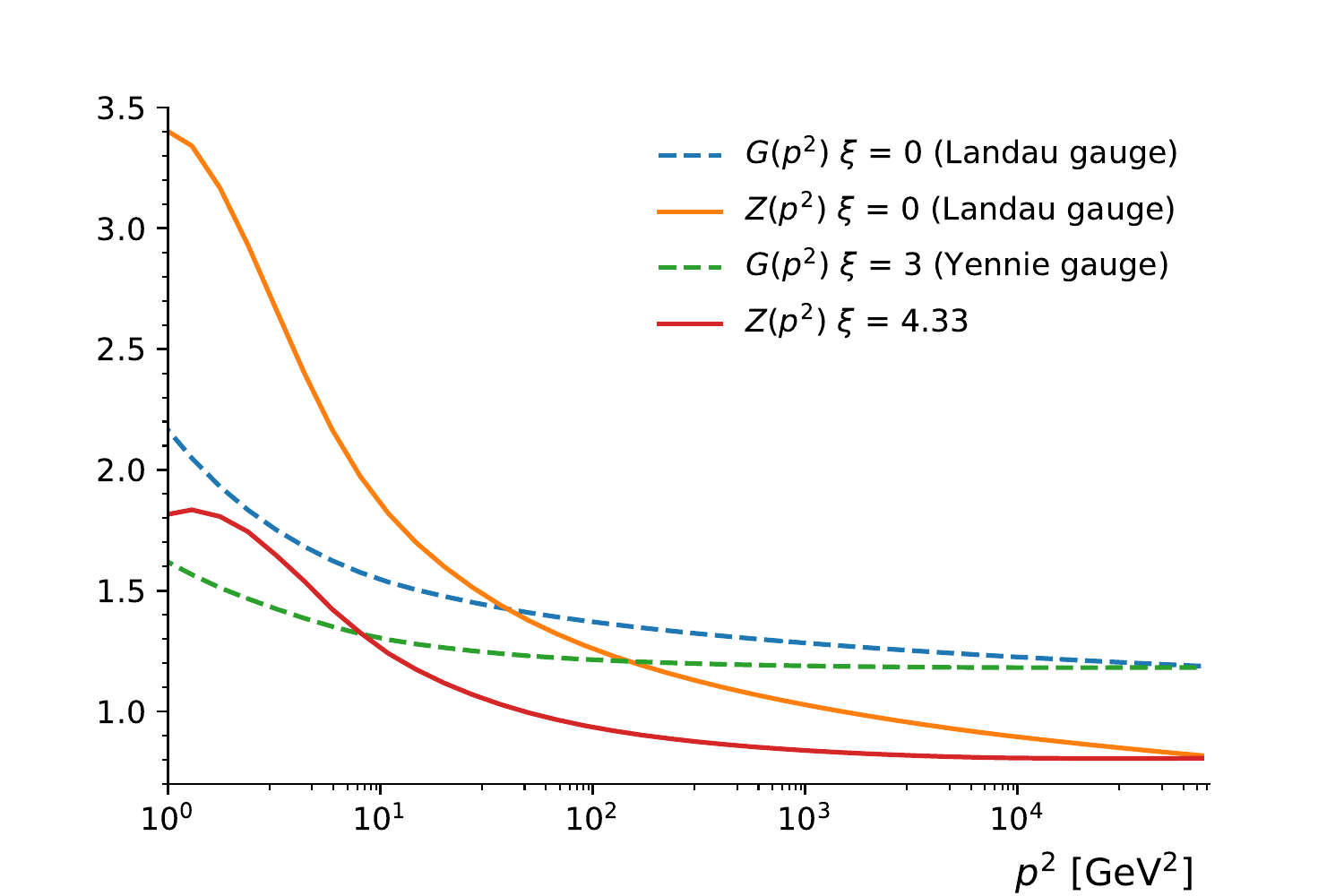}
 \caption{The UV behavior of the ghost and gluon dressing functions for $\xi=3$ and $\xi=13/3$, respectively.}
 \label{fig:res_uv}
\end{figure}

As Ansatz for the vertices we multiply their tree-level tensors with products of the ghost and gluon dressing functions with appropriate powers, 
\begin{align}\nonumber 
 \Gamma_{A^{*a}_\mu A_\nu^b c^c}(p,q,k)&=g\,f^{abc}\,H(\overline p^2)\,g_{\mu\nu}\,,\\[1ex]\nonumber 
 \Gamma_{c^{*a}c^bc^c}(p,q,k)&=-g\,f^{abc}\,H(\overline p^2)\,,\\[1ex]
 H(x)&=\frac{G(x)^{\alpha}Z(x)^{\beta}}{G(s)^{\alpha}Z(s)^{\beta}}\,,
\end{align}
with $\overline{p}^2=(p^2+q^2+k^2)/2$.
The denominator ensures that the vertex models are unity at $\overline p^2=s$.
The antifield vertices run logarithmically like the ghost-gluon vertex, as can be checked by a perturbative one-loop analysis.
The exponents are determined such that they respect this UV behavior.
As a second condition, the integrals in the NIs should be IR finite.
We make the simple Ansatz
\begin{align}
 \alpha=\alpha_0+\alpha_1\xi\,,\qquad 
 \beta=\beta_0+\beta_1\xi\,,
\end{align}
where $\alpha_i$ and $\beta_i$ are $\xi$-independent parameters.
Enforcing the conditions above, we obtain
\begin{align}
 \alpha_0=-\frac{26 \beta_0}{9}\,,\quad 
 \alpha_1=-\frac{9-4\beta_0}{6}\,,\quad 
 \beta_1=\frac{9-4\beta_0}{12}\,.
\end{align}
$\beta_0>0$ is a free parameter for which we choose for convenience  $\beta_0=1$.
A test of the sensitivity of our results on this choice as well as a discussion of the function $H(x)$ 
is provided in Appendix \ref{app:ModelDep}. It can be seen that the parameter $\beta_0$, if chosen within a reasonable range, only influences the IR and this in a quantitatively mild way.

Both the ghost and the gluon NI have the form
\begin{align}
 \frac{\partial \ln M}{\partial \xi}=K\,.
\end{align}
$K$ denotes the integrals from the skeleton expansions of the vertices.
The formal solution to this equation is
\begin{align}
 M(\xi)=M(\xi_0)e^{\int_{\xi_0}^\xi d\xi K}\,,
\end{align}
where $\xi_0$ denotes the gauge fixing parameter for a known solution. The quantity $K$ is obtained by numerically calculating the integrals which are standard one-loop two-point integrals. As a computational framework we use \textit{CrasyDSE} \cite{Huber:2011xc}. The integrals are logarithmically divergent and need to be renormalized. We do so by momentum subtraction and require that the dressing functions stay the same at the highest calculated momentum point. It should be noted that no quadratic divergences \cite{Huber:2014tva} are present.
This situation has to be contrasted with the DSE or flow equation for the gluon propagator for which spurious quadratic divergences can arise, cf. Refs.\ \cite{Cyrol:2016tym, Huber:2020keu, Aguilar:2016vin}.

\section{Results}
\label{sec:results}

\begin{figure*}[tb]
 \includegraphics[width=0.49\textwidth]{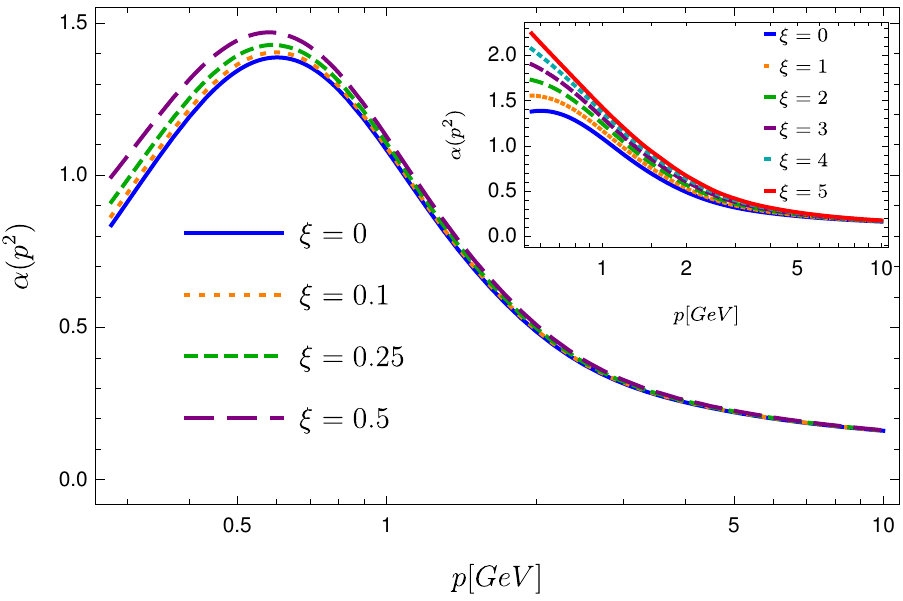}\hfill
 \includegraphics[width=0.49\textwidth]{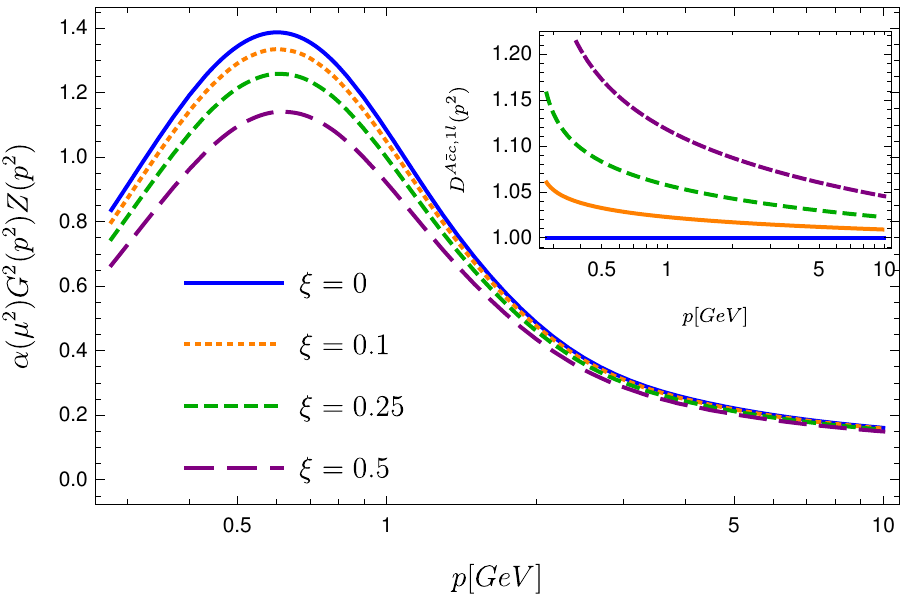}
 \caption{Left: The coupling for various values of $\xi$.
 Right: The coupling without the vertex dressing.
 The inset shows the one-loop expressions for the ghost-gluon vertex, \eref{eq:ghg1L}, used for the couplings.
 }
 \label{fig:coupling}
\end{figure*}

For the full nonperturbative solution, we solve the differential equations (\ref{eq:NIs}) up to $\xi=5$ starting from the Landau gauge. The check of the self-consistency of the UV limit is deferred to Appendix~\ref{app:UVlimit}. 

The results for the gluon dressing functions and the gluon propagators are shown in \fref{fig:res_gl}. With increasing $\xi$, the gluon propagator decreases.
This can be seen in both the maximum of the dressing function and the IR behavior of the propagator.
The difference between Landau gauge and $\xi=0.5$ is not drastic and compatible with lattice results \cite{Bicudo:2015rma}.
We find at $0$ and $1\,\text{GeV}$ that the gluon propagator goes down by $8\%$ and $7\%$, respectively, for $\xi=0.5$.
For the lattice results these ratios are given in Ref.~\cite{Bicudo:2015rma} as approximately $10\%$ and $5\%$ 
with errors of a few percentage points each.
It should be noted, though, that the IR behavior of the gluon propagator depends on the employed models for the antifield vertices.
All individual diagrams in the gluon NI are IR finite as shown in \fref{fig:res_diags} of Appendix~\ref{app:K(i)}.
This comes from the IR behavior of the anti-field vertices.
If we used bare vertices, IR divergences would arise that would qualitatively change the IR behavior of the gluon propagator.

The ghost propagator shows for $\xi>0$ the already known logarithmic IR suppression \cite{Aguilar:2015nqa,Huber:2015ria}, see \fref{fig:res_gh}.
This behavior results from diagram (a), see \fref{fig:res_diags}.
For small $\xi$, the deviation from the Landau gauge is not very large.
This agrees with lattice results which do not see a change in the ghost propagator up to $\xi=0.3$ 
and above approximately $500\,\text{MeV}$, which was the lowest accessible momentum value \cite{Cucchieri:2018leo,Cucchieri:2018doy}.
In the continuum results displayed here, evaluated down to 0.01~MeV, we do, however, see deviations from the Landau gauge behavior below 500 MeV also for small values of the gauge parameter $\xi$.
Sizeable deviations appear for higher values of the gauge fixing parameter for which also the effect on the UV behavior becomes visible.

Given the comparatively simple approximation employed for the NIs, the agreement with lattice results is very good. On the other hand, the method is very stable, and we calculated up to $\xi=5$ without encountering any problems. Indeed, we  can easily check that for $\xi=3$ and $\xi=13/3$ the correct one-loop anomalous dimensions are produced, as for these values the ghost and the gluon anomalous dimensions vanish, respectively.
This is illustrated in \fref{fig:res_uv} where the corresponding propagators are compared to the Landau gauge ones 
in the momentum region from 1 to $10^5$ GeV$^2$.

The ghost-gluon coupling is defined via the relation
\begin{align}
 \alpha(p^2)=\alpha(\mu^2)G^2(p^2)Z(p^2)[D^{A\bar c c}(p^2)]^2\,,
\end{align}
where the ghost-gluon vertex dressing $D^{A\bar c c}$ is evaluated at the symmetric point, and $\alpha(\mu^2)=g^2/(4\pi)$.
One-loop universality entails that any dependence of the coupling on the gauge fixing parameter is suppressed at high momenta.
Beyond one-loop, a dependence on $\xi$ can appear, see, e.g., Refs.~\cite{Davydychev:1998kj,Davydychev:1997vh,Chetyrkin:2000dq}.
However, we can still assess the effect of the truncation by comparing the couplings in the perturbative regime above a few GeV.
For the coupling the correct running of all quantities is important.
We thus use the one-loop resummed expression for the ghost-gluon vertex,
\begin{align}\label{eq:ghg1L}
 D^{A\bar c c,1l}(p^2)=D^{A\bar c c}(s)\left(1+\omega \ln\frac{p^2}{s}\right)^{-3\xi/22}\,,
\end{align}
where $\omega=11\,N_c\,\alpha(s)/(12\pi)G^2(s)Z(s)[D^{A\bar c c}(s)]^2$.
We show the couplings for various values of $\xi$ in \fref{fig:coupling}. Hereby, the scale $s$ is chosen as $10^5\,\text{GeV}^2$, phrased otherwise, the couplings are fixed at this value.
As can be seen, down to $\lesssim$ 10~GeV they agree even for higher values of the gauge fixing parameters.
Up to $\xi=0.5$, the agreement is good down to approximately 2~GeV.
To appreciate this agreement, we also show the coupling without the ghost-gluon vertex dressing in \fref{fig:coupling}.
The agreement is worse then and the order of magnitudes is different with $\xi=0$ having the largest coupling.
We additionally show the one-loop vertex dressing in the inset to highlight that the vertex dressing becomes sizable already for low $\xi$.
It would be interesting to test the $\xi$ dependence of related quantites like the effective charge defined in \cite{Aguilar:2009nf}.

\begin{figure}[bt]
 \includegraphics[width=0.49\textwidth]{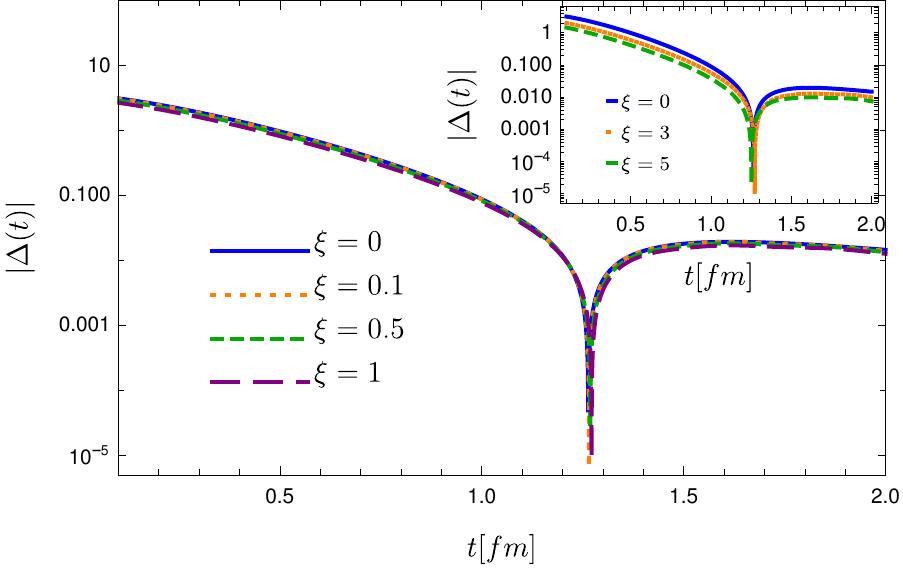}
 \caption{The absolute value of the Schwinger function for various values of the gauge fixing parameter.
 The Schwinger function is negative for $t\gtrsim1.3\,\text{fm}$.}
 \label{fig:SchwingerFunction}
\end{figure}

\begin{figure*}[tb]
\includegraphics[width=0.48\textwidth]{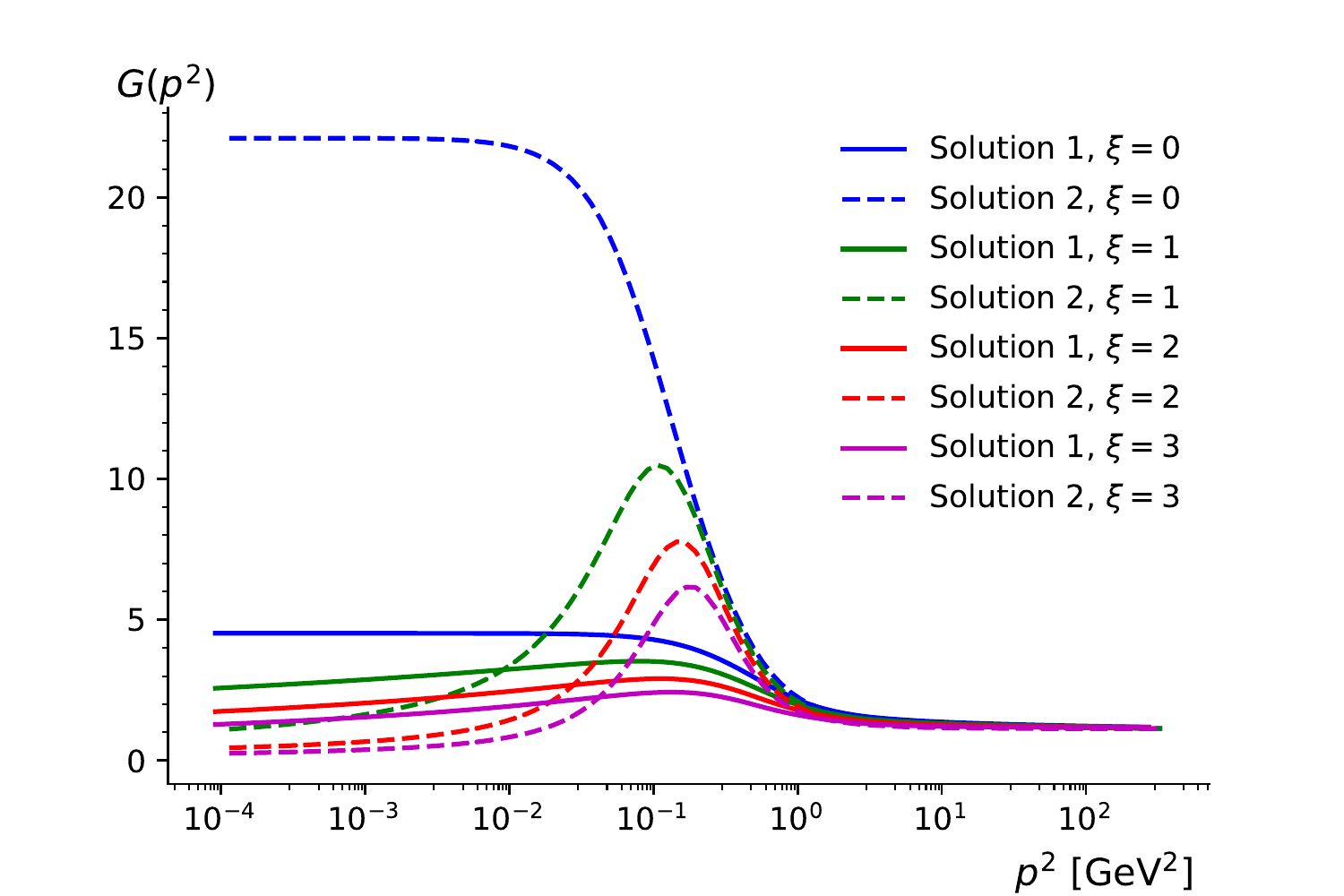}
\includegraphics[width=0.48\textwidth]{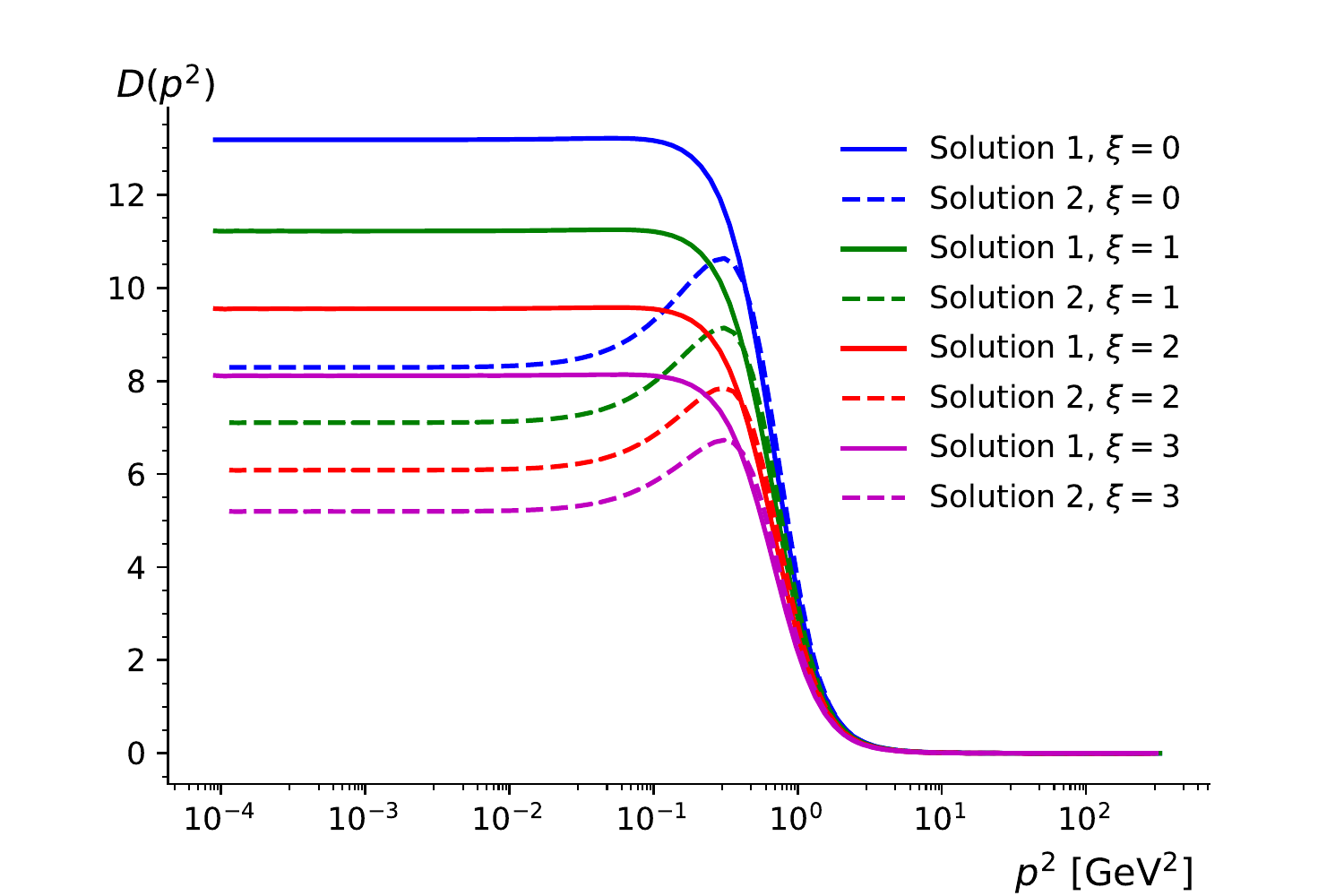}
 \caption{Ghost dressing function (left) and gluon propagator (right) for two different decoupling solutions as starting points at various value of $\xi$.}
 \label{fig:comp_sols}
\end{figure*}

\begin{figure}[tb]
\includegraphics[width=0.49\textwidth]{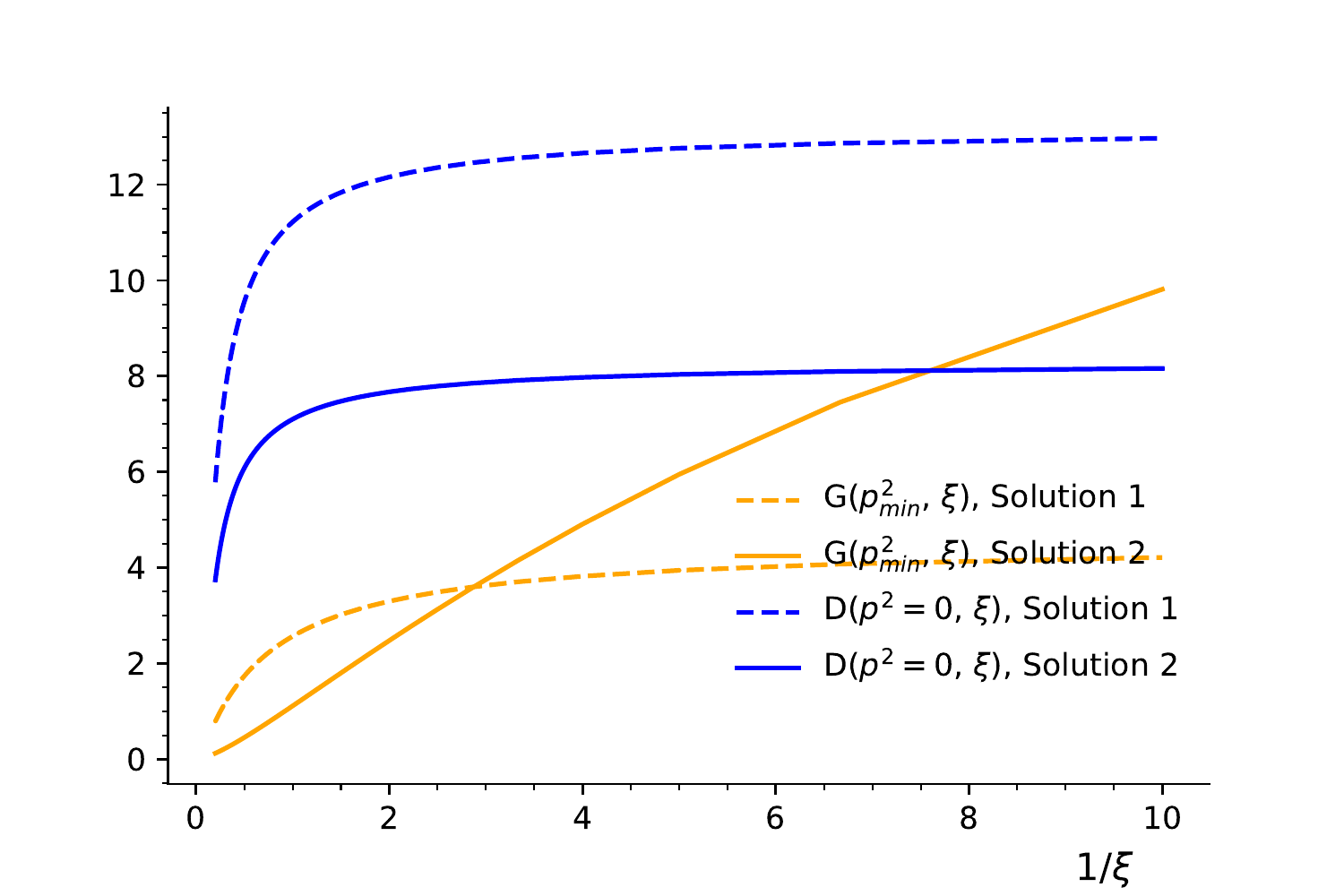}
 \caption{The gauge parameter dependence of the ghost dressing function (dashed line) and gluon propagator (solid line) at fixed momenta.
 The lowest calculated value is for $1/\xi=1/5$.}
 \label{fig:xi_dependence}
\end{figure}

\begin{figure}[tb]
\includegraphics[width=0.48\textwidth]{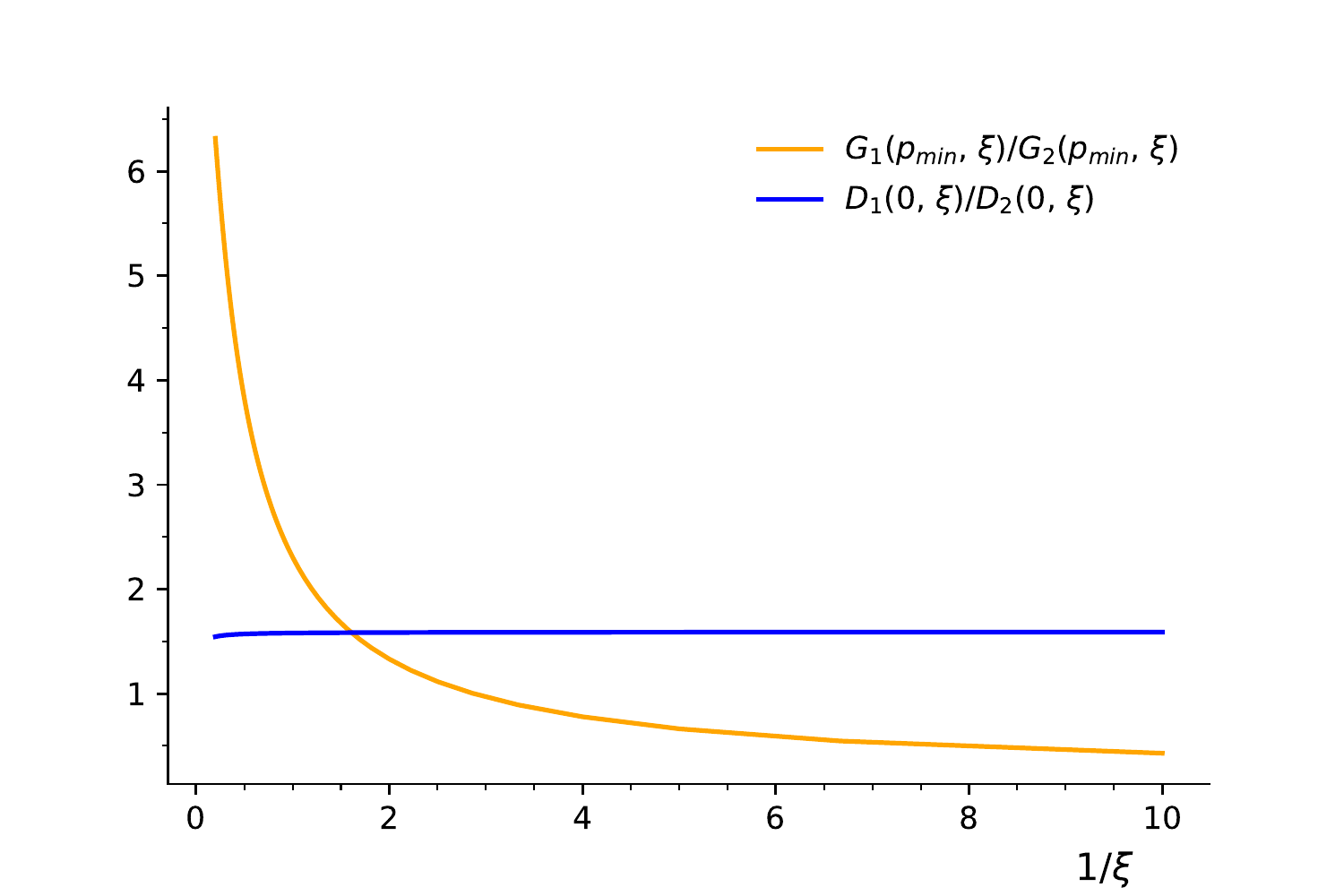}
 \caption{The ratio of the propagators for two solutions.
 For the ghost propagator the lowest calculated momentum is used, for the gluon propagator zero momentum.}
 \label{fig:xi_ratio}
\end{figure}

Another interesting quantity is the Schwinger function $\Delta(t)$ of the gluon propagator, defined as the Fourier transformation of the momentum space propagator for vanishing spatial momentum.
If the propagator violates positivity, this is reflected in the Schwinger function.
Fig.~\ref{fig:SchwingerFunction} shows the Schwinger function for various values of the gauge fixing parameter.
Up to approximately $\xi=1$ the Schwinger function barely changes.
In particular, the position of the zero crossing does not move.
Only for higher values of $\xi$ it moves slightly.
This stability is in marked contrast to the situation for the family of solutions in the Landau gauge where the position of the zero crossing moves \cite{Huber:2020keu}.
This can be understood as the existence of the zero crossing is related to the maximum of the gluon propagator.
For linear covariant gauges, we find that the position of this maximum is basically constant in $\xi$.
Different members of the family of solutions in the Landau gauge, on the other hand, exhibit different positions for the maxima \cite{Cyrol:2016tym,Huber:2020keu} and hence the Schwinger function also changes.

Finally, to explore the fate of the family of different solutions for correlation functions in the Landau gauge \cite{Boucaud:2008ji, Fischer:2008uz, Alkofer:2008jy, Maas:2009se, Maas:2011se, Sternbeck:2012mf}, we also solved the NIs using a second Landau gauge solution. In this context it should be mentioned that these different solutions may correspond to different nonperturbative infrared gauge completions of the perturbative Landau gauge as discussed in \cite{Fischer:2008uz, Maas:2009se}. If this conjecture is correct, this would correspond to a second gauge fixing direction in addition to $\xi$. Indeed, all observables in Yang-Mills theory and QCD, computed so far within this potential family of infrared completions of the Landau gauge agree within the respective error bars. A specifically relevant example in the present context of Yang-Mills theories is provided by the glueball masses, see \cite{Huber:2020ngt}. In line with the conjecture discussed above, the obtained masses did not show any deviations within errors. 

In all plots in this section we used up to here the solution that is closest to lattice results.
It is characterized by a very flat maximum of the gluon propagator, and a ghost dressing function that is relatively small in the deep infrared. 
The different Landau gauge solution used next as a starting point for the NIs is shown in \fref{fig:propsLG} in comparison
to the previously used solution.
The second solution has a pronounced maximum in the gluon propagator and shows a clear increase of the ghost dressing function at low momenta. The results for a selection of values of $\xi$ are shown in \fref{fig:comp_sols}. The typical features of the Landau gauge solution type is inherited by the $\xi>0$ ones. In particular, the gluon propagator has a shallow/pronounced maximum from which it follows immediately that it violates positivity.
Note that such a property also leads to a spectral dimension of
one in the deep IR \cite{Kern:2019nzx}. Correspondingly, if the maximum vanished, this would imply a qualitative change of the type of solution, and it is reassuring that we do not observe that. 

Since a nonzero gauge fixing parameter washes out the gauge fixing condition of the Landau gauge and thus the differences between the two solutions, it is interesting to check if the two solutions approach each other for high values of $\xi$.
To assess that, we plot the ghost dressing function and the gluon propagator at fixed momenta as a function of $\xi$ in \fref{fig:xi_dependence}.
For the ghost dressing function we see that the two solutions come closer to each other for higher values.
For the gluon propagator, on the other hand, this effect is not observed.
In both cases it seems plausible that for $\xi\rightarrow \infty$ the functions vanish.
This limit corresponds to removing the gauge fixing which in fact necessarily will eventually lead to a vanishing gluon propagator.
Phrased otherwise, we see the expected behavior based on the general properties of the linear covariant gauges.
This provides some confidence that the overall qualitative behavior of the propagators is correct for all allowable values
of the gauge parameter $0< \xi < \infty$.

The distinct relative behavior of the two solutions is clearly visible in \fref{fig:xi_ratio} which shows the ratio between the propagators at a fixed momentum point in the IR.
The ratio stays constant for the gluon propagator but depends strongly on the gauge fixing parameter for the ghost propagator.
Since the plotted ratio is $G_1/G_2$, this means that the ghost propagator with higher values in the IR for the Landau gauge decreases faster than the one with lower values.

\section{Summary}
\label{sec:summary}

We have calculated the ghost and gluon propagators of Yang-Mills theory, respectively, quenched QCD, in the linear covariant gauges for values of the gauge fixing parameter $0<\xi \leq 5$. The starting point has been results in the Landau gauge, $\xi=0$, which were obtained in a self-contained DSE calculation. As external input we employed the nonperturbative parts of the ghost-gluon and three-gluon vertices of Landau gauge for all values of $\xi$.
In addition, we used ans\"atze for the antifield vertices which contain one free parameter.
We found that the solutions are not very sensitive to variations of this parameter.

In the IR, we recover the logarithmic suppression of the ghost dressing function, predicted by earlier investigations, and find an IR finite gluon propagator. The latter result, however, happens by construction based on the antifield vertex model.
All results agree well, even quantitatively, with available lattice results.

Compared to other methods, our setup is quite stable, even at values of $\xi$ beyond the Feynman gauge. In particular, we recover the correct UV behavior for the propagators most convincingly seen by vanishing anomalous dimensions for the ghost and gluon dressing functions at $\xi=3$ (Yennie gauge) and $\xi=13/3$, respectively. We did not encounter any signs of instability up to the highest calculated value, $\xi=5$. 

While the changes of propagators and vertices are sizable, observables are $\xi$-independent. This calls for respective studies of e.g.\ glueball masses as done in \cite{Huber:2020ngt} for the Landau gauge. Such a study was beyond the scope of the present work. Instead, as a first step in this direction, we have discussed the $\xi$ dependence of the ghost-gluon coupling, \fref{fig:coupling}, and the zero crossing of the Schwinger function, \fref{fig:SchwingerFunction}. We have shown that the $\xi$ dependence of both, the coupling as well as the  
Schwinger function zero crossing, are very small up to $\xi=0.5$, which is highly nontrivial. Beyond  $\xi=0.5$, the reliability of the current approximation is successively getting worse because we do not consider the back-coupling of the $\xi$ dependence in the vertices.
Nevertheless, the observed deviations are still quite small.

We also have explored the potential family of nonperturbative infrared completions, as discussed in the Landau gauge. We have tested two different starting points and obtained two corresponding sets of solutions for $\xi>0$. The qualitative features, in particular violation of positivity, remain intact at least up to $\xi=5$. In the limit of infinite $\xi$, both propagators are in agreement with the expectation that they vanish in this limit.

In the present work we have restricted ourselves to pure Yang-Mills theory. However, the inclusion of dynamical quarks is straightforward as there are no direct quark contributions in the gluon and ghost Nielsen identities \cite{Breckenridge:1994gs}. Moreover, the Nielsen identity for the quark propagator has a similar structure as those for the other propagators and could be solved within a skeleton expansion. It would be also interesting to extend the current study to other covariant gauges like the maximally Abelian gauge. There, direct calculations are complicated due to its IR dominant two-loop diagrams \cite{Huber:2009wh,Huber:2011fw}.

\section*{Acknowledgments}

We thank Joannis Papavassiliou for discussions.
Support by the FWF (Austrian science fund) under Contract No. P27380-N27 is gratefully acknowledged. This work is supported by EMMI and the BMBF grant 05P18VHFCA. It is part of and supported by the DFG Collaborative Research Centre SFB 1225 (ISOQUANT) and the DFG under Germany's Excellence Strategy EXC - 2181/1 - 390900948 (the Heidelberg Excellence Cluster STRUCTURES).

\appendix

\begin{figure*}[tb]
\includegraphics[width=0.48\textwidth]{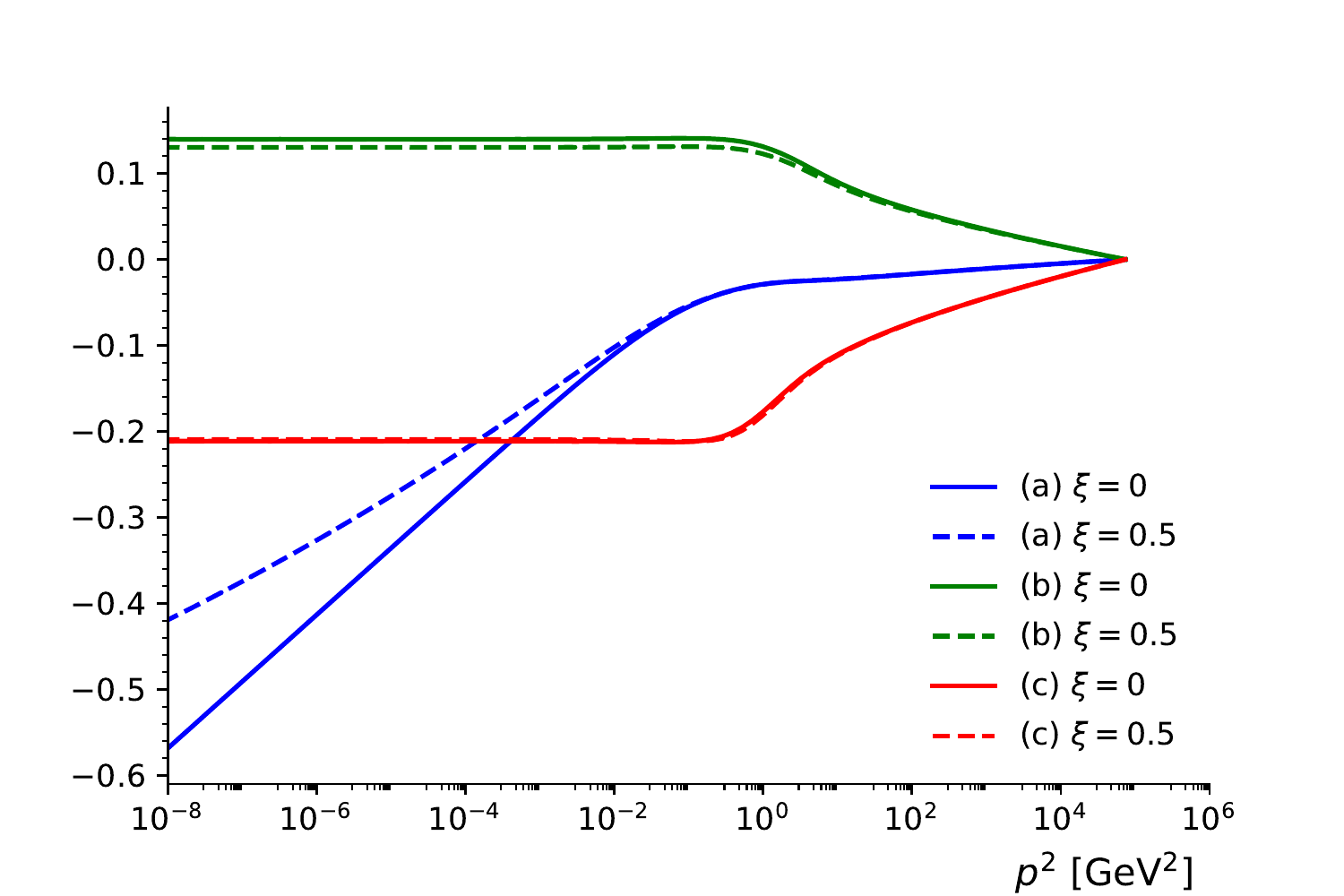}
\includegraphics[width=0.48\textwidth]{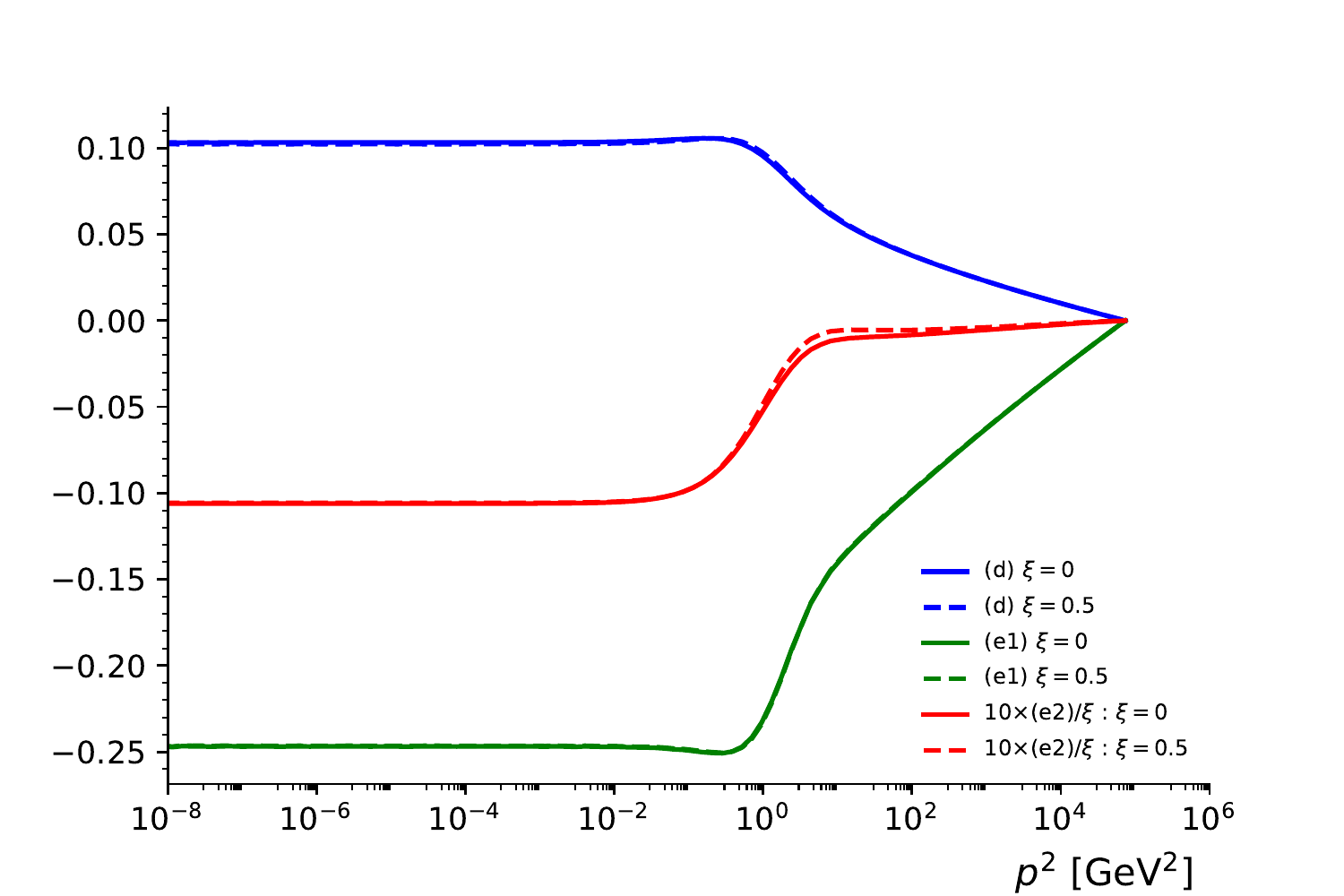}
 \caption{Contributions of individual diagrams at two values of $\xi$ for the ghost (left) and the gluon (right) NIs.}
 \label{fig:res_diags}
\end{figure*}

\begin{widetext}

\section{Diagrams in the Nielsen identities of the propagators }\label{app:K(i)}

The loops displayed in \fref{fig:skelExp} lead to the following expressions:
	\begin{subequations}
		\label{eq:Ks}
		\begin{align}
			K^{(a)} &= -\frac{N_c g^2}{2} \int \frac{d^4q}{(2\pi)^4} \frac{G(q)G(p+q)}{q^2 (p+q)^2}  D_L^{A \overline c c}(-p;p+q,-q) D_L^{A \overline c c}(q;p,-p-q),   \\[2ex]
			K^{(b)} &= -\frac{N_c g^2}{2} \int \frac{d^4q}{(2\pi)^4} \frac{G(q)Z(p+q)}{p^2 q^4 (p+q)^2} D_T^{A \overline cc}(-p-q;p,q)(p^2\,q^2-(p\cdot q)^2))C^{AAA}(-p,p+q,-q),   \\[2ex]
			K^{(c)} &= -\frac{N_c g^2}{2} \int \frac{d^4q}{(2\pi)^4} \frac{G(q)G(p+q)}{q^2 (p+q)^2}  D_L^{A \overline c c}(q;-p-q,p) \Gamma^{c^*cc}(-p,p+q,-q),   \\[2ex]
			K^{(d)} &= -\frac{N_c g^2}{3} \int \frac{d^4q}{(2\pi)^4} \frac{ G(q)G(p+q) }{p^2 q^4 (p+q)^2}(p^2\,q^2-(p\cdot q)^2)) D_T^{A\overline cc}(-p;p+q,q) \Gamma^{AA^*c}(-q,-p,p+q),    \\[2ex]\nonumber 
			K^{(e_1)} &= \frac{N_c g^2}{3} \int \frac{d^4q}{(2\pi)^4} \frac{Z(p+q)G(q)}{p^2 q^4(p+q)^4} (q^2+2\,p\cdot q)(3\,p^4+(p\cdot q)^2+2\,p^2(q^2+3\,p\cdot q)) \\[1ex]
			&\quad \times C^{AAA}(-p,p+q,-q) \Gamma^{AA^*c}(p+q,-p,-q), \\[2ex]
			K^{(e_2)} &= \frac{N_c g^2}{3}\xi \int \frac{d^4q}{(2\pi)^4} \frac{G(q)}{ q^4(p+q)^4} (p^2q^2-(p\cdot q)^2) C^{AAA}(-p,p+q,-q) \Gamma^{AA^*c}(p+q,-p,-q).
		\end{align}
	\end{subequations}
\end{widetext}
The last integral was split into two parts to disentangle the contribution from the transverse and longitudinal parts of the gluon propagator.
We used $p_\mu q_\nu r_\rho \Gamma^{AAA}_{\mu\nu\rho}(p,q,r)=0$ in several places to simplify the expressions.
E.g., due to this only the transversely projected part of the ghost-gluon vertex appears in the second diagram.
The results for the individual $K^{(i)}$ are shown in \fref{fig:res_diags} for two different values of $\xi$.

\section{Renormalization group properties of correlation functions}\label{app:RGrescalings}

Here we provide the RG rescalings of the correlation functions and couplings ${\cal O}_i$ used in the NIs, schematically 
provided in (\ref{eq:RGrescalings}). 

Since all quantities in the identities are renormalized ones, they behave under a change of the renormalization group scale $\mu\rightarrow\nu$ as follows:
\begin{align}\nonumber 
		D_{\mu\nu}(p^2,\mu^2)&=\frac{Z_3(\Lambda,\nu)}{Z_3(\Lambda,\mu)}D_{\mu\nu}(p^2,\nu^2),\\[1ex]\nonumber 
		D_G(p^2,\mu^2)&=\frac{\widetilde Z_3(\Lambda,\nu)}{\widetilde Z_3(\Lambda,\mu)}D_G(p^2,\nu^2)\,,\\[1ex]\nonumber 
		D^{AAA}(p_i^2,\mu^2)&=\frac{Z_1(\Lambda,\mu)}{Z_1(\Lambda,\nu)}D^{AAA}(p_i^2,\nu^2)\,,\\[1ex]\nonumber 
		D^{A\bar c c}(p_i^2,\mu^2)&=\frac{\widetilde Z_1(\Lambda,\mu)}{\widetilde Z_1(\Lambda,\nu)}D^{A\bar c c}(p_i^2,\nu^2)\,,\\[1ex]\nonumber 
		D^{c* c c}(p_i^2,\mu^2)&=\frac{\widetilde Z_1(\Lambda,\mu)}{\widetilde Z_1(\Lambda,\nu)}D^{c* c c}(p_i^2,\nu^2)\,,\\[1ex]\nonumber 
		D^{A*A c}(p_i^2,\mu^2)&=\frac{\widetilde Z_1(\Lambda,\mu)}{\widetilde Z_1(\Lambda,\nu)}D^{A*A c}(p_i^2,\nu^2)\,,\\[1ex]\nonumber 
		g(\mu)&=\frac{Z_g(\Lambda,\nu)}{Z_g(\Lambda,\mu)}g(\nu)\,,\\[1ex]
		\label{eq:rg_trafo_xi}
		\xi(\mu)&=\frac{Z_3(\Lambda,\nu)}{Z_3(\Lambda,\mu)}\xi(\nu)\,,
	\end{align}
where  $Z_3$, $\widetilde Z_3$, $Z_1$, $\widetilde Z_1$, and $Z_g$ are the renormalization constants for the gluon propagator, the ghost propagator, the three-gluon vertex, the ghost-gluon vertex, and the coupling which are related by the STIs
\begin{align}\label{eq:STIs}
	Z_1^2=Z_g^2 Z_3^3,\quad \widetilde Z_1^2=Z_g^2 Z_3\widetilde Z_3^2\,.
\end{align}

\section{Consistent UV limit of correlation functions}\label{app:UVlimit}

Here we discuss the self-consistency of the UV behavior of the NIs. The employed approximation is exact at the perturbative one-loop level. Consequently, on the right hand side the correct $\xi$-dependent part of the anomalous dimension must emerge. This can be seen as follows. Consider the loop integrals with bare dressing functions for large loop momenta $q$. The angle integrals can then be performed and lead to
\begin{align}\nonumber 
	(a) &\xrightarrow{UV} \frac{3 \omega}{88} \ln\left(\frac{p^2}{\mu^2}\right)\,, \\[1ex]\nonumber 
	(b) &\xrightarrow{UV} - \frac{9 \omega}{88} \ln\left(\frac{p^2}{\mu^2}\right)\,, \\[1ex]\nonumber 
	(c) &\xrightarrow{UV} \frac{3 \omega}{22} \ln\left(\frac{p^2}{\mu^2}\right)\,, \\[1ex]\nonumber 
	(d) &\xrightarrow{UV} -\frac{3\omega}{44} \ln\left(\frac{p^2}{\mu^2}\right)\,, \\[1ex]\nonumber 
	(e_1) &\xrightarrow{UV} \frac{9 \omega}{44} \ln\left(\frac{p^2}{\mu^2}\right)\,, \\[1ex]
	(e_2) &\xrightarrow{UV} 0\,.
\label{eq:asymp_behav_mom_overview}\end{align}
Only the logarithmic parts were kept and $\mu$ is a renormalization scale. Summing up the corresponding coefficients leads to 
\begin{align}
	\partial_\xi\delta=\frac{3}{44}\,,\qquad  \partial_\xi \gamma=\frac{3}{22}\,,
\end{align}
as can be checked with \tref{tab:anom_dims}. This is consistent to one-loop order with the left-hand side of the equation.
\begin{table}[h]
	\begin{tabular}{|l|c|}
		\hline
		 & Anomalous dimension \\[2mm]
		\hline \hline
		Ghost propagator & $\delta=-\frac{9-3\xi}{44} $\\[1mm]
		\hline
		Gluon propagator & $\gamma=-\frac{13-3\xi}{22}$\\[1mm]
		\hline
		Ghost-gluon vertex & $\gamma^{ghg}=-\frac{3\xi}{22}$\\[1mm]
		\hline
		Three-gluon vertex & $\gamma^{\text{3g}}=\frac{17-9\xi}{44}$\\[1mm]
		\hline
	\end{tabular}
	\caption{\label{tab:anom_dims}The one-loop anomalous dimensions of the propagators and vertices.}
\end{table}

\bigskip

\section{Dependence on the parameter $\beta_0$ }\label{app:ModelDep}

\begin{figure}[tb]
\includegraphics[width=0.48\textwidth]{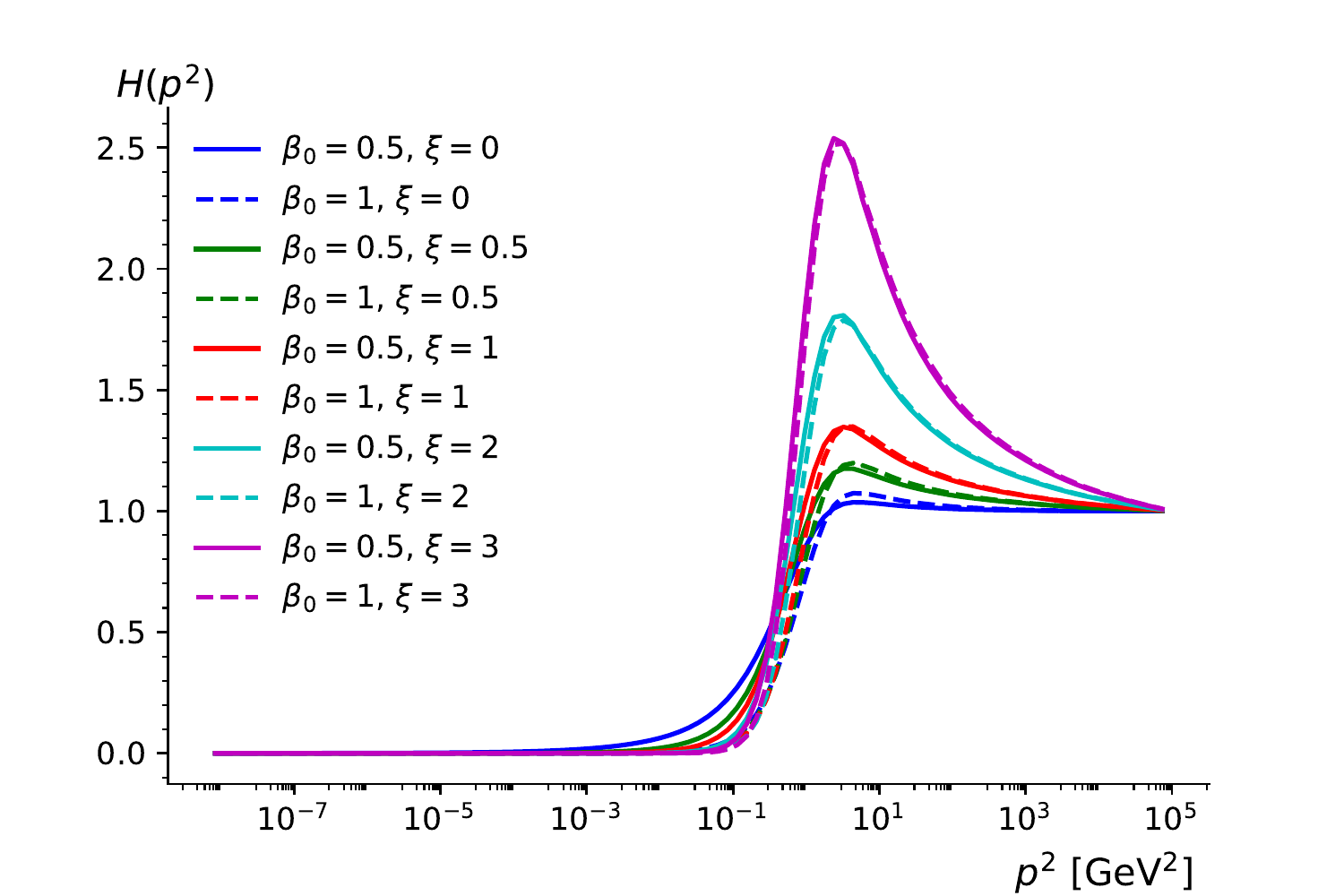}
 \caption{The model employed for the antifield vertices for two values of the parameter $\beta_0$ at different values for $\xi$.}
 \label{fig:model}
\end{figure}

The model we employ for the antifield vertices depends on one parameter $\beta_0$.
For the results shown in the main part we used $\beta_0=1$.
We tested the influence of this parameter by calculating the propagators also with $\beta_0=0.5$.
The model function $H(\bar p^2)$ for these two choices is shown in \fref{fig:model}.
In the quantitatively relevant regime around $1\,\text{GeV}$ the two parameter values lead only to small differences for all $\xi$.
The rise in the UV for larger $\xi$ comes directly from the anomalous dimension of the antifield vertices.

\begin{figure}[tb]
\includegraphics[width=0.48\textwidth]{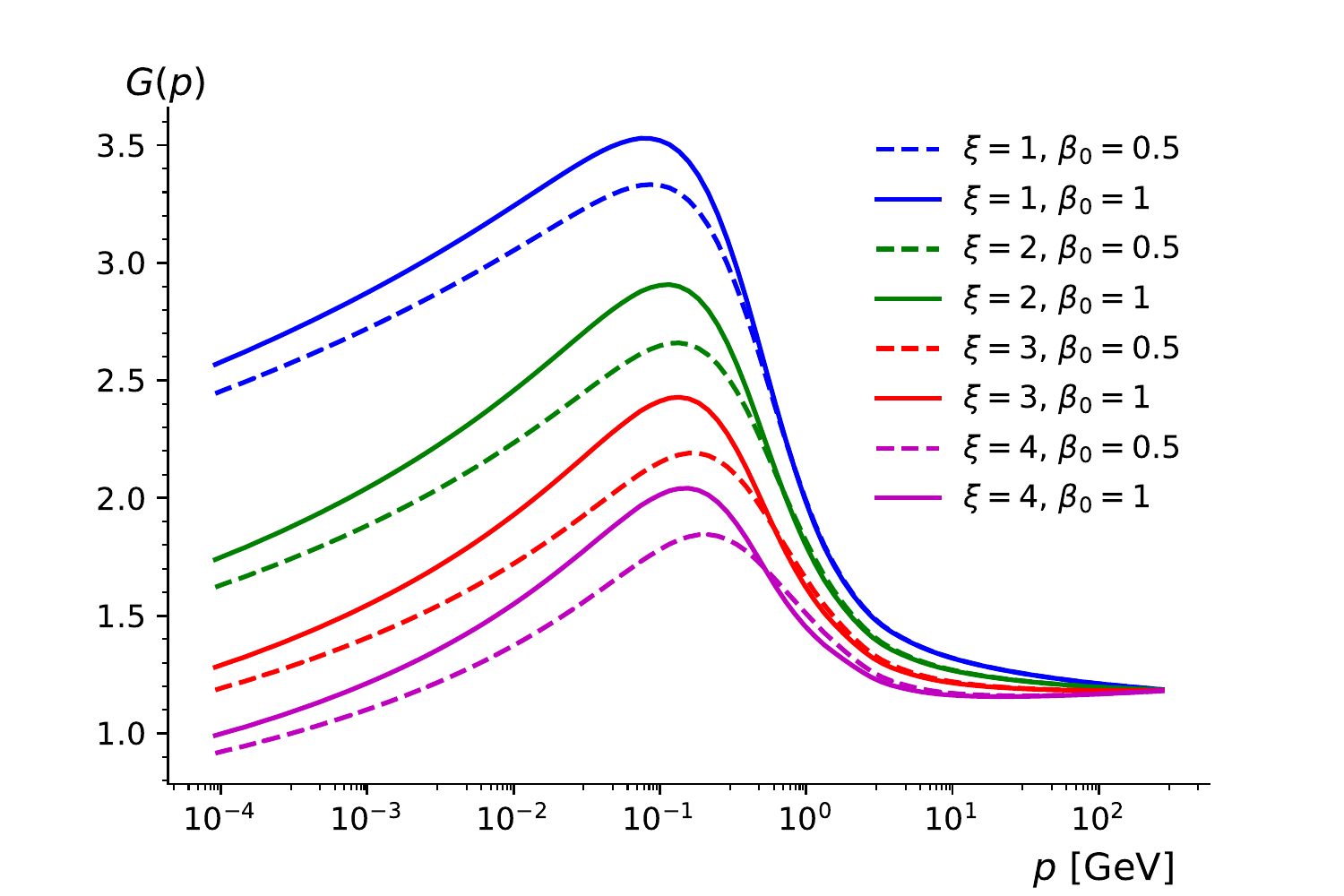}
\includegraphics[width=0.48\textwidth]{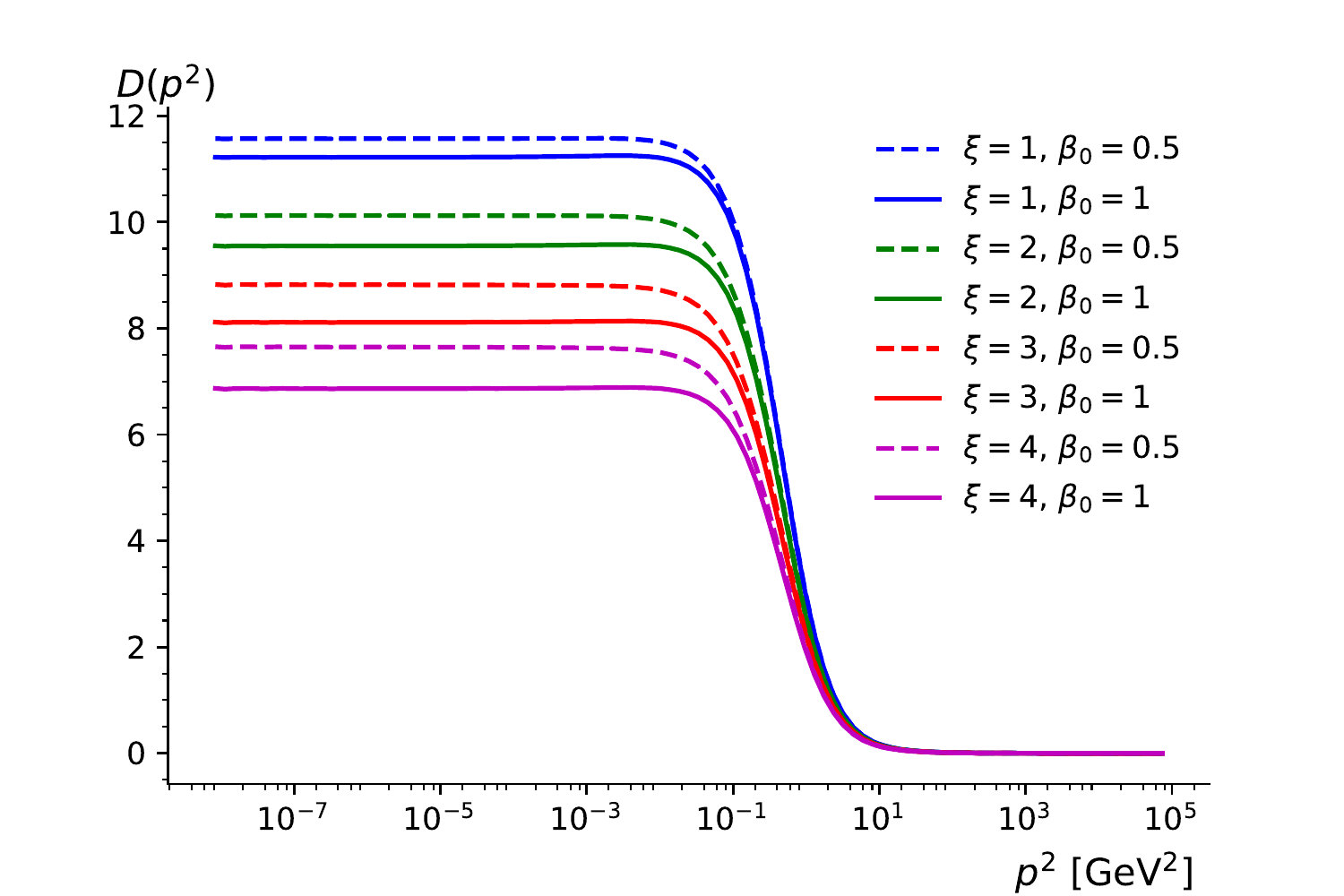}
 \caption{Ghost dressing function (top) and gluon propagator (bottom) for $\beta_0=0.5$ and $1$ at various value of $\xi$.}
 \label{fig:comp_beta0}
\end{figure}

The propagators obtained from $\beta_0=0.5$ are compared to the ones from $\beta_0=1$ in \fref{fig:comp_beta0}.
We can see that changing $\beta_0$ affects basically only the IR.
Only for $\xi>4$ small effects in the ghost dressing function are seen also in the midmomentum regime.
We thus conclude that within the present approximation scheme the dependence on the model for the antifield vertices is of minor importance and only seen quantitatively for low momenta.

\FloatBarrier

\bibliographystyle{utphys_mod}
\bibliography{literature_LinCov_Nielsen}

\end{document}